\documentclass[11pt]{article}
\usepackage[dvipdfmx]{color}
\usepackage{graphicx}
\usepackage{xcolor}
  \usepackage{amsthm,amsfonts}
  \usepackage{amsmath}
\usepackage{mathabx}
\usepackage{slashed}
\usepackage{ulem}
\usepackage{hyperref}
\usepackage{cleveref}
\usepackage{cite}

\newcommand{\bea}   {\begin{eqnarray}}
\newcommand{\eea}   {\end{eqnarray}}
\def\zzg{${\mathbb Z}_2\times{\mathbb Z}_2$-graded }

\begin{document}
\renewcommand{\thefootnote}{\fnsymbol{footnote}}

\thispagestyle{empty}

\title{Volichenko-type metasymmetry of braided Majorana qubits}
\author{ Francesco Toppan\thanks{{E-mail: {\it toppan@cbpf.br}}}
\\
\\
}
\maketitle

{\centerline{
{\it CBPF, Rua Dr. Xavier Sigaud 150, Urca,}}\centerline{\it{
cep 22290-180, Rio de Janeiro (RJ), Brazil.}}
~\\
\maketitle

\begin{abstract}
~\\
This paper presents  different mathematical structures  connected with the parastatistics of braided Majorana qubits and clarifies their role; in particular, ``mixed-bracket" Heisenberg-Lie algebras are introduced. These algebras belong to a more general framework than the {\it Volichenko algebras} defined in 1990 by Leites-Serganova as {\it  metasymmetries} which do not respect even/odd gradings and lead to mixed brackets interpolating ordinary commutators and anticommutators.\\
In a previous paper braided ${\mathbb Z}_2$-graded Majorana qubits were first-quantized within a graded Hopf algebra framework endowed with a braided tensor product. The resulting system admits truncations at 
roots of unity and realizes, for a given integer $s=2,3,4,\ldots$, an interpolation between ordinary Majorana fermions (recovered at $s=2$) and bosons (recovered in the $s\rightarrow \infty$ limit); it implements a parastatistics where at most $s-1$ indistinguishable particles are accommodated in a multi-particle sector.\\
The structures discussed in this work are:\\
- the quantum group interpretation of the roots of unity truncations recovered from a (superselected) set of reps of the quantum superalgebra ${\cal U}_q({\mathfrak{osp}}(1|2))$;\\
- the reconstruction, via suitable intertwining operators, of the braided tensor products as ordinary tensor products
(in a minimal representation, the $N$-particle sector of the braided Majorana qubits is described by $2^N\times 2^N$ matrices);\\
- the introduction of mixed brackets for the braided creation/annihilation operators which define generalized Heisenberg-Lie algebras;\\
- the $s\rightarrow \infty$ untruncated limit of the mixed-bracket Heisenberg-Lie algebras producing parafermionic oscillators;\\
-   ({\it meta})symmetries of ordinary differential equations given by matrix Schr\"{o}dinger equations in $0+1$ dimension induced by the braided creation/annihilation operators;\\
- in the special case of a third root of unity truncation, a nonminimal realization of the intertwining operators defines the system
as a ternary algebra.

\end{abstract}
\vfill
\rightline{CBPF-NF-002/24}
\newpage

\section{Introduction}

Emergent Majorana fermions (${\mathbb Z}_2$-graded Majorana qubits) have been intensively investigated in the light of the Kitaev's proposal \cite{kit} (see also\cite{{nssfds}, {brki}, {kau}}) to use them for encoding topological quantum computations which offer topological protection from quantum decoherence.\par
It was shown in \cite{topqubits} that the braiding of ${\mathbb Z}_2$-graded Majorana qubits can be realized within a graded Hopf algebra endowed with a braided tensor product (this approach was based on the framework proposed in \cite{maj}). In the present paper the First Quantization of ${\mathbb Z}_2$-graded braided Majorana qubits is shown to be connected and expressed by several different constructions. We briefly sketch them.
\par
At first this paper solves a question left unanswered in \cite{{topqubits},{topscipost}}: how to directly relate the observed roots-of-unity truncations of the multi-particle braided spectra to quantum group reps data; it is shown that the truncations
are also recovered from (superselected) representations
of the quantum group ${\cal U}_q({\mathfrak{osp}}(1|2))$.\par
On the technical side, the realization of the braided tensor products in terms of intertwining operators acting on ordinary tensor products paves the way for the further constructions presented in the paper.\par
The next relevant topic regards the parastatistics induced by the braided Majorana qubits. An
$s^{th}$- root of unity truncation realizes, for a given integer $s=2,3,4,\ldots$, an interpolation between ordinary Majorana fermions (recovered at $s=2$) and the untruncated case recovered in the $s\rightarrow \infty$ limit. For the
$s$-th root at most $s-1$ indistinguishable particles are accommodated in a multi-particle sector. \par
These ``level-$s$ parastatistics" are reproduced via the introduction of generalized Heisenberg-Lie algebras defined by ``mixed brackets" which interpolate ordinary commutators (denoted as ``$[.,.]$") and anticommutators (denoted as ``\{.,.\}").  Given two operators $X,Y$ the $(.,.)_{\vartheta_{XY}}$ mixed bracket is defined as  
\bea
(X,Y)_{\vartheta_{XY}} &:=& i \sin(\vartheta_{XY}) \cdot [X,Y] + \cos(\vartheta_{XY})\cdot \{X,Y\},\nonumber
\eea
where the $\vartheta_{XY}$ angle is consistently determined.\par
 These new mixed-bracket Heisenberg-Lie algebras broadly fit into the notion of ``symmetries wider than supersymmetry" presented by Leites and Serganova in \cite{lese1} (see also \cite{lese2}); this notion concerns the existence of statistics-changing maps which do not preserve  the ${\mathbb Z}_2$-grading of ordinary Lie superalgebras.\par
The Leites-Serganova construction is focused on ``Volichenko algebras" which act as {\it metasymmetries}  which do not respect even/odd gradings. A Volichenko algebra satisfies a ``metaabelianess condition" ({\it metaabelianess} means that, for any $X,Y,Z$ triple of operators, the ordinary $[[X,Y],Z]=0$ commutators are vanishing).
 The introduction of mixed brackets which interpolate ordinary commutators/anticommutators is also a natural ingredient of the Leites-Serganova construction, see e.g. formula (1.3.2) of reference \cite{lese2}. \par
The mixed-bracket Heisenberg-Lie algebras are not Volichenko algebras since they violate the metaabelianess
condition for some choices of the $X,Y,Z$ operators. Nevertheless, they satisfy an alternative identity involving the mixed brackets: for any $X,Y,Z$ triple of operators, the relation $(X,(Y,Z))=0$ is satisfied. This identity can be referred to as ``metaabelianess with respect to the mixed brackets".\par
Furthermore, the notion of {\it metasymmetry} can be applied to the mixed-bracket Heisenberg-Lie algebras. Indeed, it is shown in Section {\bf 10} that these mixed-bracket algebras close dynamical (meta)symmetries for given systems of Matrix Ordinary Differential Equations. \par
It is pointed out that in the $s\rightarrow \infty$ untruncated limit the mixed-bracket Heisenberg-Lie algebras reduce to (anti)commutators defining a set of  parafermionic oscillators.\par
In Appendix {\bf A} we clarify that the mixed-bracket parastatistics induced by the generalized Heisenberg-Lie algebras {\it are not} reproduced by another type of mixed-bracket construction, the quonic oscillator interpolation introduced by Greenberg \cite{gre2} and by Mohapatra \cite{moh}.  \par
In Appendix {\bf B} a nonminimal realization of the intertwining operators for the third root of unity allows to make contact with the ternary algebras entering the applications of ternary physics (for ternary algebras see \cite{akl} and references therein).\par
We postpone to the Conclusions a more detailed assessment of the found connections relating multi-particle braided ${\mathbb Z}_2$-graded Majorana qubits and the various afore-mentioned mathematical structures (quantum group reps at roots of unity, truncated parastatistics and the untruncated $s\rightarrow \infty$ limit, metasymmetries induced by mixed-bracket algebras, etc.).\\
~\par
The scheme of the paper is the following:\\
~\\
- In Section {\bf 2} we revisit the \cite{topqubits} construction of the multi-particle ${\mathbb Z}_2$-graded braided Majorana qubits obtained from a graded Hopf algebra endowed with a compatible braided tensor product;\\
- In Section {\bf 3} we present different suitable parametrizations of the truncations of the spectra at roots of unity;\\
- In Section {\bf 4} we recover the roots-of-unity truncations from a quantum group perspective, pointing out that it is given by (superselected) reps of the quantum supergroup ${\cal U}_q({\mathfrak{osp}}(1|2))$;\\
- In Section {\bf 5} the braiding of Majorana qubits is realized via intertwining operators acting on ordinary (i.e., not braided) tensor products;\\
- In Section {\bf 6} the indistiguishability of identical particles is recovered as a superselection;\\
- Section {\bf 7} presents, following Leites-Serganova's papers, a sketchy introduction to the notion of the Volichenko algebras
and their related metasymmetries;\\
- Section {\bf 8} presents the generalized ``mixed-bracket" Heisenberg-Lie algebras which interpolate bosons/fermions and reproduce the multi-particle sectors of the graded Majorana qubits;\\
- In Section {\bf 9} it is shown that the $s\rightarrow \infty$ limit of the mixed-bracket Heisenberg-Lie algebras reduce to (anti)commutators describing a set of parafermionic oscillators;\\
- Section {\bf 10} presents the mixed-bracket Heisenberg-Lie algebras as dynamical metasymmetries of an ordinary differential Matrix Schr\"odinger equation in $0+1$ dimensions;\\
-  In the Conclusions we give comments about the link of the presented constructions with parastatistics and discuss future perspectives;\\
- Appendix {\bf A} clarifies the difference between generalized mixed-bracket Heisenberg-Lie algebras and the 
parastatistics induced by quons;\\
- In Appendix {\bf B} a nonminimal realization of the intertwining operators for the third root of unity is related to ternary algebras.

\section{The braided Majorana qubits revisited}

It was shown in \cite{topqubits} that ${\mathbb Z}_2$-graded Majorana qubits can be braided within a First Quantization formalism. We revisit in this Section the main ingredients of this  construction which makes use of a graded Hopf algebra endowed with a braided tensor product. \par
~\par
The starting point of \cite{topqubits} is the ${\mathfrak{gl}}(1|1)$ superalgebra. 
A single Majorana fermion describes a ${\mathbb Z}_2$-graded qubit which defines a bosonic  vacuum state $|0\rangle$ and a fermionic excited state $ |1\rangle$:
{\small{\bea\label{qubit01}
|0\rangle = \left(\begin{array}{c} 1\\0\end{array}\right) , &\quad& |1\rangle =\left(\begin{array}{c} 0\\1\end{array}\right) .
\eea}}
The following operators, acting on the ${\mathbb Z}_2$-graded qubit, close ${\mathfrak{gl}}(1|1)$:
{\small{\bea\label{4op}
&\alpha =\left(\begin{array}{cc} 1&0\\0&0\end{array}\right),\quad ~\beta =\left(\begin{array}{cc} 0&1\\0&0\end{array}\right),\quad ~ \gamma =\left(\begin{array}{cc} 0&0\\1&0\end{array}\right),\quad ~\delta =\left(\begin{array}{cc} 0&0\\0&1\end{array}\right).&
\eea}}
The (anti)commutators are
\bea\label{anticomm}
&\relax [\alpha,\beta]=\beta, \qquad [\alpha,\gamma]=-\gamma,\qquad  [\alpha,\delta]=0,\qquad [\delta,\beta]=-\beta,\qquad  [\delta,\gamma]=\gamma,&\nonumber\\
&~\quad\{\beta,\beta\}=\{\gamma,\gamma\}=0,~\quad\qquad\qquad \{\beta,\gamma\}=\alpha+\delta.&
\eea
The diagonal operators $\alpha,\delta$ are even, while $\beta,\gamma$ are odd ($\gamma$ is the fermionic
creation operator).\par
The excited state $ |1\rangle$ is a Majorana fermion which coincides with its own antiparticle. The diagonal operator $\delta$ can be assumed to be the Hamiltonian $H_1$ (therefore, $H_1\equiv \delta$).\par 
The Universal Enveloping Algebra ${\cal U}\equiv {\cal U}({\mathfrak{gl}}(1|1))$, which is a graded Hopf algebra, allows to introduce the multi-particle ${\mathbb Z}_2$-graded qubits. A non-trivial brading of the Majorana qubits is realized by introducing, following the \cite{maj} prescription, a braided tensor product (here denoted as ``$\otimes_{br}$") which is compatible with the  ${\mathfrak{gl}}(1|1)$ superalgebra. This means in particular that, for a nonvanishing complex parameter $t\in {\mathbb C}^\ast ={\mathbb C}\backslash \{0\}$, the creation operator $\gamma$  can be assumed to satisfy the relation
\bea \label{braidingamma}
({\mathbb I}_2\otimes_{br} \gamma)\cdot (\gamma\otimes_{br} {\mathbb I}_2) &=& B_t\cdot (\gamma\otimes_{br} {\mathbb I}_2)\cdot ({\mathbb I}_2\otimes_{br} \gamma) \equiv B_t\cdot (\gamma\otimes_{br}\gamma),
\eea
where the ``$\cdot$" symbol denotes the standard matrix multiplication; here and in the following ``${\mathbb I}_n$"denotes the $n\times n $ identity matrix.
The $t$-dependent $4\times 4$ constant matrix $B_t$ is invertible for $t\neq 0$. It can be assumed to be the $R$-matrix of the Alexander-Conway polynomial in the linear crystal rep on exterior algebra
\cite{kasa}; it is related, see \cite{rsw}, to the Burau representation of the braid group. The matrix $B_t$ is given by {\footnotesize{\bea\label{btmatrix}
B_t&=&\left(\begin{array}{cccc} 1&0&0&0\\0&1-t&t&0\\0&1&0&0\\0&0&0&-t\end{array}\right).
\eea
}}
The ${\otimes}_{br}$ braided tensor product  defined in (\ref{braidingamma}) satisfies the required \cite{maj} compatibility  conditions; this is a consequence of the $B_t$ matrix satisfying the braid relation
\bea\label{braidedrel}
(B_t\otimes {\mathbb I}_2)\cdot ({\mathbb I}_2\otimes B_t)\cdot 
(B_t\otimes {\mathbb I}_2) &=& ({\mathbb I}_2\otimes B_t) \cdot
(B_t\otimes {\mathbb I}_2)\cdot ({\mathbb I}_2\otimes B_t).
\eea
\subsection{The construction of the multi-particle states}}

Let ${\cal H}$ be  the single-particle Hilbert space spanned by $|0\rangle ,|1\rangle$ entering (\ref{qubit01}). The $N$-particle Hilbert space ${\cal H}_N$ is a subset of $N$ tensor products of the single-particle Hilbert spaces:
 \bea\label{subspace}
{\cal H}_{N}&\subset &{\cal H}^{\otimes N}.
\eea
The $N$-particle vacuum state $|0\rangle_{N}$ is defined as
\bea
\qquad |0\rangle_N &=& |0\rangle \otimes \ldots \otimes |0\rangle\qquad \qquad (\textrm{$N$ times}),
\eea
while the $N$-particle excited states are obtained by repeatedly applying on $|0\rangle_N$ the coproducts of the creation operator
$\gamma$. \par
The coproduct  map $\Delta$ sends
\bea\label{coproductbr}
\Delta~:~{\cal U}\rightarrow {\cal U} \otimes_{br}  {\cal U},&& \Delta^{(N+1)} := (\Delta\otimes_{br} id)\Delta^{(N)}=(id\otimes_{br} \Delta)\Delta^{(N)}\in {\cal U}^{\otimes_{br} N}.
\eea
The second relation expresses the coassociativity of the coproduct.\par
The further property
\bea\label{uaub}
   \Delta(U_AU_B)&=&\Delta(U_A)\Delta(U_B) \qquad{\textrm{for any ~$U_A,U_B\in {\cal U}$}}
\eea
implies that the action on any given $U\in {\cal U}({\mathfrak gl}(1|1))$ is recovered from the action of the coproduct on the
Hopf algebra unit ${\bf 1}$ and on the primitive elements $\zeta\in {\mathfrak{gl}(1|1)}$; they are respectively given by
\bea\label{deltaidg}
   \Delta({\bf 1})={\bf 1}\otimes_{br}{\bf 1}, &\quad & 
   \Delta({ \zeta})={\bf 1}\otimes_{br}{\zeta}+\zeta\otimes_{br} {\bf 1}.
\eea
In our conventions a hat denotes the evaluation of the coproduct on a given representation $R$ of the Universal Enveloping Algebra ${\cal U}$, so that
\bea\label{hat}
&{\textrm{for}}\quad R: {\cal U}\rightarrow V,\qquad {\widehat \Delta}:= \Delta|_R\in End (V\otimes V),\qquad 
 {\textrm{with}}\quad  {\widehat{ \Delta({U})}}\in V\otimes V&
\eea
and, similarly, ${\widehat{ \Delta^{(N)}(U)}}\in V\otimes \ldots \otimes V$ taken $N+1$ times. For the case under consideration the vector space $V$ is given by ${\cal H}$.
It deserves being pointed out that the presence of a hat indicates an ordinary $m\times m$ matrix; this is why ordinary tensor products appear  in formula (\ref{hat}), while braided tensor products appear in (\ref{coproductbr}).\par
The $N$-particle Hilbert space ${\cal H}_N$ is spanned by
the normalized vectors $|n\rangle_{t,N}$, where the integer $n$ labels the $n$-th excited state; we have 
\bea\label{spanning}
|n\rangle_{t,N} &\propto& {\widehat{\left({{ {\Delta^{(N-1)}}(\gamma)}}\right)^{n}}} |0\rangle_N, \qquad {\textrm{for $n=0,1,2,\ldots$.}}
\eea
The $N$-particle Hamiltonian $H_N$ is expressed in terms of $H_1=\delta=diag(0,1)$ as
\bea\label{nparticlehamiltonians}
H_N &:=& {{ {\widehat{ \Delta^{(N-1)}}}(H_1)}}.
\eea
At the lowest orders the $N=2,3$-particle Hamiltonians are
\bea\label{h2and3}
H_{2} &=&H_1\otimes {\mathbb I}_2+{\mathbb I}_2\otimes H_1= diag(0,1,1,2),\nonumber\\
H_{3} &=&H_1\otimes{\mathbb I}_2\otimes{\mathbb I}_2 +{\mathbb I}_2\otimes H_1\otimes{\mathbb I}_2+{\mathbb I}_2\otimes{\mathbb I}_2\otimes H_1= diag(0,1,1,2,1,2,2,3).
\eea
The $N=2,3$-particle coproducts of the creation operator $\gamma$ are
\bea\label{gamma2and3}
  {{ \Delta (\gamma )}}&=&\gamma \otimes_{br}{\mathbb I}_2+ {\mathbb I}_2\otimes_{br}\gamma, \nonumber\\
 {{ \Delta^{(2)}(\gamma)}}&=&\gamma\otimes_{br}{\mathbb I}_2\otimes_{br}{\mathbb I}_2+{\mathbb I}_2\otimes_{br} \gamma\otimes_{br}{\mathbb I}_2+ {\mathbb I}_2\otimes_{br}{\mathbb I}_2\otimes_{br}\gamma .
\eea
For later convenience it is useful to split the $N=2,3$-particle coproducts of the creation operator into the following
building blocks:
\bea\label{twothree}
{\textrm{For~ $N=2$}}&:& A_1^\dagger := \gamma \otimes_{br}{\mathbb I}_2,\qquad \quad ~ A_2^\dagger := {\mathbb I}_2\otimes_{br}\gamma. \nonumber\\
{\textrm{For~ $N=3$}}&:& B_1^\dagger := \gamma\otimes_{br}{\mathbb I}_2\otimes_{br}{\mathbb I}_2,~~B_2^\dagger := {\mathbb I}_2\otimes_{br}\gamma\otimes_{br}{\mathbb I}_2,~~B_3^\dagger := {\mathbb I}_2\otimes_{br}{\mathbb I}_2\otimes_{br}\gamma .
\eea 
Equation (\ref{braidingamma}) implies in particular that
\bea\label{commuting}
A_2^\dagger\cdot A_1^\dagger &=& -tA_1^\dagger\cdot A_2^\dagger \qquad~~ {\textrm{and}}\nonumber\\
B_2^\dagger\cdot B_1^\dagger &=& -tB_1^\dagger\cdot B_2^\dagger,\qquad B_3^\dagger\cdot B_1^\dagger ~=-tB_1^\dagger\cdot B_3^\dagger,\qquad B_3^\dagger\cdot B_2^\dagger ~= -tB_2^\dagger\cdot B_3^\dagger.
\eea
By taking into account the nilpotency of $\gamma$ we get
\bea\label{nilpotency}
{\gamma^2=0} &\Rightarrow & \quad{\textrm{ $(A_i^\dagger)^2=0$~~ for $i=1,2$~~ and~~ $(B_j^\dagger)^2=0$ ~~for $j=1,2,3$.}}
\eea
It follows that the coproduct of the $2$-particle second-excited state is given by
\bea\label{braidedcoproductsquare}
 ({{ \Delta (\gamma )}})^2&=&(1-t) A_1^\dagger\cdot A_2^\dagger.
\eea 
The splitting into building blocks is extended to the coproducts of the $N$-particle sectors.
We can set, for $k=1,\ldots, N$,
\bea\label{general}
{A_{k}^{\dagger}}_{;N}&:=&\underbrace{ {\mathbb I_2}\otimes_{br} \ldots\otimes_{br} {\mathbb I}_2}_\text{$k-1$ times}\otimes_{br}\gamma\otimes_{br} \underbrace{{\mathbb I}_2\otimes_{br} \ldots \otimes_{br} {\mathbb I}_2}_\text{$N-k$ times},\nonumber\\
 {{ \Delta^{(N-1)}(\gamma)}}&=& \sum_{k=1}^{N-1} {A_{k}^{\dagger}}_{;N}.
\eea
Formulas (\ref{twothree},\ref{commuting}) are recovered from the positions $A_i^\dagger \equiv {A_{i}^{\dagger}}_{;2}$ and $B_j^\dagger\equiv {A_{j}^{\dagger}}_{;3}$.\par
In the next Section we summarize the results concerning the multi-particle spectra of the braided Majorana qubits
and the truncations recovered at certain $t$ roots of unity.
~\par
~\par

\section{The truncations at roots of unity}

In this Section we summarize the results concerning the multi-particle energy spectra of the
braided ${\mathbb Z}_2$-graded Majorana qubits. Since the derivation of these results has been presented in \cite{topqubits}, it will not be repeated here. We stress in particular the role of the truncations which are  recovered at certain roots of unity of the braiding parameter $t$ entering (\ref{braidingamma},\ref{btmatrix}). Different useful parametrizations of $t$,
which will be used in the following, are introduced.\par
~\par
The $k=1,\ldots, N$ building blocks ${A_{k}^{\dagger}}_{;N}$ entering (\ref{general}) satisfy the braiding relations
\bea\label{generalcommuting}
{A_{k'}^{\dagger}}_{;N}\cdot {A_{k}^{\dagger}}_{;N}&= &-t {A_{k}^{\dagger}}_{;N}\cdot {A_{k'}^{\dagger}}_{;N}\qquad {\textrm{ for ~$k<k'$.}}
\eea
In consequence of that the operators $ {\widehat{\left({{ {\Delta^{(N-1)}}(\gamma)}}\right)^{n}}}$  (entering the (\ref{spanning}) vectors
$|n\rangle_{t,N}$ spanning the $N$-particle Hilbert space ${\cal H}_N$) are proportional to a factor
$f_n(t)$, expressed by the $t$-dependent polynomials $b_k(t)$:
\bea\label{polynomials}
f_n(t)&=&  \prod_{k=1}^n b_k(t),\qquad {\textrm{where ~ $b_k(t)=\sum_{j=0}^{k-1} (-t)^j$.}}
\eea 
A nonvanishing vector $|n\rangle_{t,N}$ is an eigenvector of the  $N$-particle Hamiltonian $H_N$ with energy eigenvalue $E_{n;t,N}=n$:
\bea
H_N|n\rangle_{t,N} &=& E_{n;t,N} |n\rangle_{t,N}, \qquad {\textrm{where ~ $E_{n;t,N}=n$.}}
\eea
Therefore, the zeros of the $f_n(t)$ polynomial factors (which are induced by the zeros of the $b_k(t)$ polynomials defined in (\ref{polynomials})), produce truncations at the corresponding energy level spectra.\par
It is easily shown that the zeros of the $b_k(t)$ polynomials lie in the $|t|=1$ circle and are given by {\it certain} roots of unity. The {\it untruncated} spectrum is recovered for all other values of $t$, both the nonvanishing complex parameter $t$ satisfying $|t|\neq 1$ and  the generic values  of $t$ belonging to the $|t|=1$ circle which are not roots of the $b_k(t)$ polynomials. Without loss of generality,
the physics  of the multi-particle braided Majorana qubits (both truncated and untruncated sectors) can be derived by restricting $t$ to belong to the $|t|=1$ circle. \par
A first parametrization of the unit circle, see \cite{topscipost}, is introduced by setting 
\bea\label{fparametrization}
&t=e^{i\pi f}, \qquad {\textrm{for real values ~ $f\in [0, 2[$.}}&
\eea
A relevant notion is that of ``level-$s$" root of unity for $s=2,3,4,\ldots $. It is defined as follows.\\
The $b_k(t)$ polynomial is of order $k-1$ (we have $b_1(t)=1$, $b_2(t)=1-t$, $b_3(t) =1-t+t^2$, $\ldots$).
Its $k-1$ roots are roots of unity. A level-$s$ root of unity $t_s$ corresponds to a solution of the equation
\bea
b_s(t_s)&=&0
\eea
which {\it is not} a root of any previous $b_k(t)$ polynomial for $k<s$; namely
\bea
b_k(t_s) &\neq & 0 \qquad {\textrm{for $k=2,3,\ldots, s-1$.}}
\eea
Denoted as $L_s$ the set of level-$s$ roots of unity $t_s^{(j)}$ we have, at the lowest orders, the following representatives:
\bea\label{tchoices}
L_2 &:& t_2^{(1)}=1, \qquad \qquad\qquad\qquad~{\textrm{implied by~~ $b_2(t)=1-t=0$,}}\nonumber\\
L_3 &:& t_3^{(1)}= e^{\frac{i\pi}{3}},~~ t_3^{(2)}= e^{\frac{5i\pi}{3}},\qquad {\textrm{implied by~~ $b_3(t)=1-t+t^2=0$,}}\nonumber\\
L_4 &:& t_4^{(1)}= e^{\frac{i\pi}{2}},~~ t_4^{(2)}= e^{\frac{3i\pi}{2}},\qquad {\textrm{implied by~~ $b_4(t)=1-t+t^2-t^3=0$.}}
\eea
One should note that $t=1$, which is the third root of $b_4(t)=0$, is not a level-$4$ root of unity since it is already a solution of $b_2(t)=0$.\par
The $t=-1$ root of unity {\it is not a root} of any $b_n(t)$ polynomial since, for $n=1,2,3,\ldots$,
\bea
&b_n(t)_{|_{t=-1}}=n\neq 0.&
\eea
With a slight abuse of language we can say that $t=-1$ is a level-$\infty$ root, denoted as $L_\infty$.\par
In terms of $f$ introduced in (\ref{fparametrization}), the following fractions correspond to the first cases of level-$s$ roots:
\bea
&L_\infty: f_{\infty} = 1;~~\quad
L_2 : f_2= 0;~~\quad
L_3: f_3 =\frac{1}{3},~\frac{5}{3} ;~~\quad
L_4:f_4 = \frac{1}{2},~\frac{3}{2};~~\quad
L_5:f_5 = \frac{1}{5},~\frac{3}{5},~\frac{7}{5},~\frac{9}{5}.&
\nonumber\\&&
\eea

The physical significance of a level-$s$ root of unity is that the corresponding braided multi-particle Hilbert space can accommodate at most $s-1$ braided Majorana particles.\par
 The special point $t=1$, which is the level-$2$ root of unity, gives the  ordinary total antisymmetrization of the fermionic wavefunctions.  The level-$2$ root of unity $t=1$ encodes the Pauli exclusion principle of ordinary fermions.
\par
One should note that the physics does not depend on the specific value of $t$, but only on its root of unity level. A generic $t$ which does not coincide with a root of unity produces the same untruncated spectrum as the $L_\infty$ root of unity $t=-1$.\par~
\par
The following $N$-particle energy spectra are derived.  The energy levels are given by integer numbers and are not degenerate. The two big classes are:\\
~\par{{
{{{\bf Class a, truncated $L_s$ level:} the $N$-particle energy eigenvalues $E$ are
\bea\label{truncatedenergy}
E &=& 0,1,\ldots, N\qquad \quad~{\textrm{for}}\quad N<s,\nonumber\\
E &=& 0,1,\ldots, s-1 \qquad {\textrm{for}} \quad N\geq s;
\eea
in this case a plateau is reached at the maximal energy level $s-1$; this is the maximal number of braided Majorana fermions that can be accommodated in a multi-particle Hilbert space;}}\par
~\par
{{{\bf Class b, untruncated ($t=-1) ~~L_\infty$ level:} the $N$-particle energy eigenvalues $E$ are
\bea\label{genericenergy}
E &=& 0,1,\ldots, N\qquad {\textrm{for any given}}\quad N;
\eea
in this case there is no plateau and the maximal energy eigenvalues grow linearly with $N$.}}
}}
\par
~\par
A parametrization alternative to (\ref{fparametrization}) consists in expressing $t$, belonging to the $|t|=1$ unit circle, as
\bea\label{gparametrization}
&t=-e^{2i\pi g}, \qquad {\textrm{for real values ~ $g\in [0, 1[$.}}&
\eea
The connection with the (\ref{fparametrization}) parametrization of \cite{topscipost} is given by
\bea
f= -2g-1~~~mod~~2, &\quad& g= -\frac{1}{2}(f+1)~~~mod~~1.
\eea
The (\ref{gparametrization}) parametrization simplifies the analysis of the level-$s$ roots of unity; $L_s$ and the $L_\infty$ untruncated case are respectively specified by the following $g$ values
\bea
L_s&:& {\textrm{recovered from }} ~~~ g=\frac{r}{s}\qquad{\textrm{with $r,s$ mutually prime integers,}}\nonumber\\
L_\infty&:&{\textrm{recovered from}} ~~~~ g=0.
\eea
At the first orders the $g$ values are
\bea
&L_\infty: ~g_{\infty} = 0;\quad ~
L_2:  ~g_2 = \frac{1}{2};\quad ~
L_3: ~g_3 =\frac{1}{3},~\frac{2}{3} ;\quad ~
L_4:~ g_4 = \frac{1}{4},~\frac{3}{4};\quad ~
L_5: ~g_5 = \frac{1}{5},~\frac{2}{5},~\frac{3}{5},~\frac{4}{5}.&\nonumber\\
\eea
\par
Since the physics only depends on the $s$ level and not on a given particular representative, without loss of generality 
we can set $r=1$ and express the inequivalent $s$ levels as
\bea\label{gls}
L_s &:&  {\textrm{recovered from}} ~~ ~ g_s=\frac{1}{s}.
\eea
The $g=0$ case of the untruncated $L_\infty$ level is recovered in the limit
\bea
g_\infty &=& \lim_{s\rightarrow\infty} g_s =0.
\eea
The $t$ roots of unity which correspond to (\ref{gls}) are
\bea\label{ts}
t_s &=& e^{\pi i (\frac{2}{s}-1)}.
\eea

\section{A quantum group derivation of the truncations}

The results of \cite{topqubits} (see also the discussion in \cite{topscipost}) led to an open question. The truncations of the multi-particle spectra at roots of unity are reminiscent of the well-known special features of the quantum groups
representations at roots of unity, see \cite{lus} and \cite{dck}. On the other hand the approach of \cite{topqubits}
(which is based on \cite{maj}) does not directly use quantum group data. The compatible braided tensor product is applied to the $\mathfrak{gl}(1|1)$ superalgebra, not its quantum counterpart. \par
The construction which allows to recover braided multi-particle quantizations of Majorana qubits from quantum group reps is presented here. \par
It should be mentioned that the naive expectation that one should directly work with the quantum superalgebra ${\cal U}_q({\mathfrak{gl}}(1|1))$  is not viable (this quantum superalgebra and its representations have been investigated in \cite{{kul},{skd},{vir},{rsw},{sar}}). The reason is simple. The creation operator $\gamma$ entering (\ref{4op}) is nilpotent and the same is true for its ${\cal U}_q({\mathfrak{gl}}(1|1))$ quantum group counterpart.  Due to the homomorphism of the coproduct, we get the nilpotent quantum group expression
$\Delta_q(\gamma)^2=\Delta_q(\gamma)\cdot \Delta_q(\gamma)=\Delta_q(\gamma^2)=0$ for the ${\cal U}_q({\mathfrak{gl}}(1|1))$ coproduct. On the other hand, a nonvanishing $(\Delta(\gamma))^2\neq 0$ coproduct induced by the braiding, see formula (\ref{braidedcoproductsquare}), is essential to produce the multi-particle spectra of the braided Majorana qubits. \par
Clearly, some other construction has to be done.\par
The solution is found by working  within the quantum superalgebra ${\cal U}_q({\mathfrak {osp}}(1|2))$, inducing the multi-particle states
by applying its coproduct to a specific representation and, furthermore, implementing a consistent superselection of the energy spectra. These steps allow to recover the multi-particle spectra of the braided Majorana qubits.

\subsection{The ${\mathfrak{osp}}(1|2)$ superalgebra and its quantum extension}

We introduce here, following \cite{kure} (see also \cite{sal}) the quantum superalgebra ${\cal U}_q({\mathfrak{osp}}(1|2))$ as a deformation of the ${\mathfrak{osp}}(1|2)$ Lie superalgebra.  The complex deformation parameter is denoted, as in \cite{kure}, with  $\eta$ while the \cite{dic} conventions for the ${\mathfrak{osp}}(1|2)$ generators are adopted. We can identify $q$ as $q=e^\eta$.\par
The $5$-generator ${\mathfrak{osp}}(1|2)$ Lie superalgebra is defined by the nonvanishing (anti)commutators
\bea
\relax ~\qquad ~\qquad [H, F_\pm] =\pm \frac{1}{2}F_\pm, &\quad& [H,E_\pm] = \pm E_\pm,\nonumber\\
\relax [E_+,E_-] = 2H, &\quad& [E_\pm, F_\mp] =- F_\pm,\nonumber\\
\{F_+,F_-\}=\frac{1}{2} H, &\quad& \{F_\pm, F_\pm \} =\pm \frac{1}{2} E_\pm.
\eea
The two odd generators $F_\pm$ are the fermionic simple roots, while the three even generators $H, E_\pm$ define
the ${\mathfrak{sl}}(2)$ subalgebra; $H$ is the Cartan generator. \par
The second-order Casimir invariant $C_2$ is given by 
\bea
C_2 &=& H^2+\frac{1}{2}(E_+E_-+E_-E_+)-(F_+F_--F_-F_+);
\eea
$C_2$ commutes with any generator of ${\mathfrak{osp}}(1|2)$, so that  $[C_2,z]=0$ for any $ z=H, F_\pm, E_\pm$.  \par
The ${\mathfrak{osp}}(1|2)$ superalgebra possesses the graded anti-involution $\ast$, given by
\bea\label{ast}
&H^\ast = H,\quad E_\pm^\ast = E_\pm, \quad F_\pm^\ast = \pm F_\mp,&\nonumber\\& \quad {\textrm{so that }} (z^\ast)^\ast = (-1)^{\varepsilon_z} z, {\textrm{ ~with ~}} \varepsilon_H,\varepsilon_{E_\pm}=0 {\textrm{~and~}} \varepsilon_{F_\pm}=1.&
\eea 
The quantum superalgebra ${\cal U}_q ({\mathfrak{osp}}(1|2))$ is generated, see \cite{kure}, by the three elements $H, F_\pm$ satisfying, in terms of the complex parameter $\eta\neq 0$, the (anti)commutation relations
\bea\label{quantumalgebra}
\relax [H, F_\pm ]_\eta&=& \pm \frac{1}{2} F_\pm,\nonumber\\
\{F_+, F_-\}_\eta &=& \frac{\sinh (\eta H)}{\sinh(2\eta)}=\frac{e^{\eta H}-e^{-\eta H}}{e^{2\eta}-e^{-2\eta}}.
\eea
In the limit when $\eta$ goes to zero one recovers the ordinary ${\mathfrak{osp}}(1|2)$ anticommutator $\{F_+,F_-\}$:
\bea
\lim_{\eta\rightarrow 0}\{F_+,F_-\}_\eta &=&\{F_+,F_-\} =\frac{1}{2}H.
\eea
The quantum superalgebra ${\cal U}_q({\mathfrak{osp}}(1|2))$ has the structure of a graded Hopf superalgebra where, in particular, the following relations for the coproduct hold:
\bea
\Delta (H) &=& H\otimes {\bf 1}+{\bf 1}\otimes H,\nonumber\\
\Delta (F_\pm) &=& F_\pm \otimes e^{\frac{\eta}{2}H}+e^{-\frac{\eta}{2}H}\otimes F_\pm.
\eea
The tensor product is ${\mathbb Z}_2$-graded and declared to satisfy, for any set of elements, the relation
\bea
(a\otimes b)\cdot(c\otimes d) &=& (-1)^{\varepsilon_b\varepsilon_c}  ac\otimes bd \qquad {\textrm{for ~}} \varepsilon_{b,c}=0,1,
\eea
where the $\varepsilon$ grading of the generators is given in (\ref{ast}).\par
Due to the coassociativity, the $\Delta^{(2)}(F_+)$ coproduct which belongs to the tensor product of three spaces is
given by 
\bea\label{threespaces}
\Delta^{(2)}(F_+)&=& F_+ \otimes e^{\frac{\eta}{2}H}\otimes e^{\frac{\eta}{2}H}+e^{-\frac{\eta}{2}H}\otimes F_+\otimes e^{\frac{\eta}{2}H} + e^{-\frac{\eta}{2}H}\otimes e^{-\frac{\eta}{2}H}\otimes F_+.
\eea 
A single-particle Hilbert space ${\cal H}_\eta$ can be expressed as a lowest-weight representation of \\${\cal U}_q({\mathfrak{osp}}(1|2))$, defined by the Fock vacuum $|0\rangle_\eta$ such that
\bea
H|0\rangle_\eta &=&\lambda |0\rangle_\eta,\nonumber\\
F_-|0\rangle_\eta &=&0,
\eea
where $\lambda$ is a given ``vacuum energy". \par
The Hilbert space ${\cal H}_\eta$ is spanned by the (possibly infinite) series of vectors $|n\rangle_\eta$, given by
\bea
|n\rangle_\eta &=& F_+^n|0\rangle_\eta, \qquad {\textrm{where}} \quad n=0,1,2,3, \ldots .
\eea 
It follows from (\ref{quantumalgebra}) that a non-vanishing vector $|n\rangle_\eta$ is an eigenvector of $H$ with $\lambda+\frac{n}{2}$ eigenvalue:
\bea
H|n\rangle_\eta &=& (\lambda+\frac{n}{2})|n\rangle_\eta.
\eea
The representations of the quantum group ${\cal U}_q({\mathfrak{osp}}(1|2))$ and the conditions on their finiteness (which, in our conventions, depend on the pair of values $\lambda, \eta$) have been investigated in \cite{{kure},{sal},{acns}}. It is not necessary to present these conditions here since we have a very special purpose; we are interested in introducing the 
multi-particle Hilbert spaces induced by ${\cal U}_q({\mathfrak{osp}}(1|2))$ which reproduce the multi-particle braided Majorana
qubits Hilbert spaces. This implies that the single-particle Hilbert spaces have to be spanned by two vectors, $|0\rangle$ and $|1\rangle$ with respective energy eigenvalues $0$ and $1$. This implies at first that we have to set
\bea
\lambda =0 &{\textrm{and}}& H_1=2H \quad\textrm{to be the normalized single-particle Hamiltonian.}
\eea  
Since no finite two-dimensional Hilbert space is recovered from ${\cal U}_q({\mathfrak{osp}}(1|2))$, a suitable two-dimensional finite Hilbert space ${\cal H}_\eta^{(1)}$ is constructed from ${\cal H}_\eta$ by applying a projection. \par
Let us therefore introduce the $P^2=P$ projector $P=diag(1,1,0,0,0\ldots)$, defined as
\bea\label{proj}
&P|0\rangle_\eta =|0\rangle_\eta, ~~~P|1\rangle _\eta = |1\rangle_\eta, ~~~~{\textrm{while}}~~  P|n\rangle_\eta =0 ~~~{\textrm{for}} ~~~ n\geq 2.&
\eea
We can consistently set the two-dimensional, single-particle Hilbert space being given by
\bea
&{\cal H}_\eta^{(1)}\subset {\cal H}_\eta,~~{\textrm{spanned by the $P$ eigenvectors with $+1$ eigenvalue, so that $|0\rangle_\eta, |1\rangle_\eta \in {\cal H}_\eta^{(1)}$.}}&\nonumber\\&&
\eea 
The projected single-particle Hamiltonian is expressed as $PH_1P$.\par
~\par
In the multi-particle sectors,  the two-particle Hilbert space ${\cal H}_\eta^{(2)}$ is spanned by the vectors
\bea
&(P\otimes P)\cdot {\widehat{\Delta(F_+^n)}} \cdot (|0\rangle_\eta\otimes|0\rangle_\eta)\in {\cal H}_\eta^{(2)}&
\eea
(as usual in our conventions, the hat on the coproduct symbol indicates that it is evaluated in the given representation).
It is easily shown that, due to the $P$ projection, the integer $n$ can at most take the values $n=0,1,2$. \par
The $(N+1)$-particle Hilbert space ${\cal H}_\eta^{(N+1)}$ is spanned by the vectors
\bea
&(P\otimes P\otimes \ldots \otimes P)\cdot {\widehat{\Delta^{(N)}(F_+^n)}} \cdot (|0\rangle_\eta\otimes|0\rangle_\eta\otimes\ldots \otimes |0\rangle_\eta),&
\eea
where the ${\widehat{\Delta^{(N)}}}$ coproduct acts on the tensor product of $N+1$ spaces.
 Due to the $P$ projection, a simple combinatorics shows that $n$ is restricted, for generic values of the deformation parameter $\eta$, to be given by $n=0,1,2,\ldots, N+1$. 

\subsection{Recovering the roots of unity truncations}

The roots of unity truncations for the projected $N$-particle ${\cal H}_\eta^{(N)}$ Hilbert spaces are recovered by mimicking the construction presented in Section {\bf 2} for the braided Majorana qubits. For this reason we introduce the ``building blocks" of the $N$-particle creation operators ${\widehat{\Delta^{(N-1)}(F_+)}}$. Since no confusion will arise, we use the same symbols as the ones previously introduced in (\ref{twothree}). There are two reasons for that: an unnecessarily burdening of the notation is avoided and, furthermore, it is easier to make the connection with the relevant formulas,
such as (\ref{commuting}), needed for the derivation of the truncations. \par
The new definition of the building blocks is
\bea\label{quantumtwothree}
{\textrm{For~ $N=2$}}&:& A_1^\dagger := F_+ \otimes e^{\frac{\eta}{2}H},\qquad\quad  A_2^\dagger := e^{-\frac{\eta}{2}H}\otimes F_+.\nonumber\\
{\textrm{For~ $N=3$}}&:&B_1^\dagger :=F_+ \otimes e^{\frac{\eta}{2}H}\otimes e^{\frac{\eta}{2}H}, B_2^\dagger := e^{-\frac{\eta}{2}H}\otimes F_+\otimes e^{\frac{\eta}{2}H}, B_3^\dagger := e^{-\frac{\eta}{2}H}\otimes e^{-\frac{\eta}{2}H}\otimes F_+.\nonumber\\&&
\eea
Extending the new definition to the $N$-particle building blocks 
${A_{k}^{\dagger}}_{;N}$ such as those entering (\ref{general}) is immediate.\par
Contrary to their (\ref{twothree}) counterparts, the (\ref{quantumtwothree}) building blocks $A_i^\dagger, B_j^\dagger$ are not nilpotent: $(A_i^\dagger)^2, (B_j^\dagger)^2\neq 0$. On the other hand, the contributions of their squares to the multi-particle sectors is killed by applying the projector $P$ defined in (\ref{proj}). \par
By taking into account that  
$
e^{u H} F_+ e^{-u H} = e^{\frac{u}{2}}F_+
$ for a complex parameter $u$, the analogous of the (\ref{commuting}) formulas now read
\bea\label{commuting2}
A_2^\dagger\cdot A_1^\dagger &=& -e^{-\frac{\eta}{2}}A_1^\dagger\cdot A_2^\dagger \qquad~~ {\textrm{and}}\nonumber\\
B_2^\dagger\cdot B_1^\dagger &=& -e^{-\frac{\eta}{2}}B_1^\dagger\cdot B_2^\dagger,\qquad B_3^\dagger\cdot B_1^\dagger ~=-e^{-\frac{\eta}{2}}B_1^\dagger\cdot B_3^\dagger,\qquad B_3^\dagger\cdot B_2^\dagger ~= -e^{-\frac{\eta}{2}}B_2^\dagger\cdot B_3^\dagger,\nonumber\\
\eea
with the identification
\bea\label{tandeta}
t&=& e^{-\frac{\eta}{2}}.
\eea~\par
For the $2$-particle sector, the $t=1$ truncation of the $n=2$ energy level is recovered from 
\bea
{\widehat{\Delta(F_+^2)}} =(1-t) A_1^\dagger A_2^\dagger + (\ldots )&\Rightarrow &{\widehat{\Delta(F_+^2)}}_{|_{t=1}}\equiv 0,
\eea
where $(\ldots)$ denotes the extra-terms killed by the $P$ projection. We therefore reobtain the level-$2$ root of unity.\par
~\par
A straightforward computation for the $3$-particle sector shows that the contribution to the $n=3$ energy level is expressed by the formula
\bea
{\widehat{\Delta(F_+^3)}} &=& -t^{-\frac{3}{2}}(t^3-2t^2+2t-1)(F_+ e^{-\eta H}\otimes F_+\otimes F_+e^{\eta H})+({\ldots)};
\eea
once more, $({\ldots})$ denotes the extra-terms killed by the $P$ projection.\par
The truncation of the $3$-particle $n=3$ energy level is implied by
\bea
{\widehat{\Delta(F_+^3)}} \equiv 0 &\Rightarrow& t^3-2t^2+2t-1 =0.
\eea
Besides the $t=1$ level-$2$ solution, the two level-$3$ roots solving the above cubic equations are $t=e^{\pm \frac{\pi i}{3}}$.
\par
~\par
In terms of the (\ref{gparametrization})  parametrization $t=-e^{2\pi i g}$  we can set from (\ref{tandeta}):
\bea
\eta&=& -2\pi i (2g-1).
\eea
By identifying the level-$s$ roots as  $g_s= \frac{1}{s}$ from (\ref{gls}) and $t_s=e^{\pi i(\frac{2}{s}-1)}$ from  (\ref{ts}) we get, in particular,
\bea
g=\frac{1}{2} ~~&\Rightarrow& ~~\eta_2=0, \quad \qquad ~ ~t_2 =1,\nonumber\\
g=\frac{1}{3} ~~&\Rightarrow&~~ \eta_3=\frac{2\pi i}{3}, \quad \quad ~ t_3 =e^{-\frac{\pi i}{3}},\nonumber\\
g=0 ~~&\Rightarrow&~~ \eta_\infty=2\pi i, \quad\quad~  t_\infty =- 1.
\eea
where the last formula corresponds to the untruncated case. 
\par
~\par
The level-$3$ root corresponds to a third root of unity of the quantum group derivation via the identifications
\bea
(-t_3)^3 =1 &{\textrm{and}}& q_3= e^{\eta_3}\quad \Rightarrow \quad q_3^3=1.
\eea

The extension of the analysis to generic $n$ energy levels for the $N$-particle sectors is immediate. It follows from the equivalence of the building blocks commutation relations such as (\ref{quantumtwothree}) for $N=2,3$, with the (\ref{twothree}) commutation relations under the identification 
$t= e^{-\frac{\eta}{2}}$.

\section{Braided tensor products via intertwining operators}

In this Section we present an alternative description (with respect to the one  based on \cite{topqubits} and discussed in Section {\bf 2}) of the multi-particle quantization of the braided Majorana qubits. Following \cite{maj}, the $\otimes_{br}$ tensor product which enters the Section {\bf 2} construction is consistently {\it declared} to be braided. On the other hand, in many applications it would be useful to realize this construction via an ordinary tensor product $\otimes$ (the braiding being introduced via suitably defined intertwining operators). This approach is presented here.
Specifically, we replace the (\ref{twothree},\ref{general}) building blocks entering the multi-particle creation operators with
new building blocks constructed in terms of the intertwining operators. The key issue is to reproduce the nilpotency conditions (\ref{nilpotency}) and the (\ref{commuting},\ref{generalcommuting}) commutation relations.  These relations are necessary and sufficient to obtain the multi-particle spectra of the braided Majorana qubits.  \par
As already done in the previous Section, in order not to burden the notation and make easier the contact with the results of Section {\bf 2}, the same symbols will be used for the new building blocks; no confusion will arise because it is clearly specified which is which.\par
We introduce here a minimal realization of the intertwining operators which are realized by $2\times 2$ constant matrices (a non-minimal realization which is applied to the third root of unity truncation is presented in Appendix {\bf B}).
\par
~\par
The braided tensor product $\otimes_{br}$ applied to the $\gamma=${\footnotesize{$\left(\begin{array}{cc} 0&0\\1&0\end{array}\right)$}} creation operator introduced in (\ref{4op}) satisfies the relation (\ref{braidingamma}) which, with the (\ref{btmatrix}) position for $B_t$, gives
\bea\label{braidedtensorproduct}
&({\mathbb I}_2~\otimes_{br} \gamma ) \cdot (\gamma~\otimes_{br}{\mathbb I}_2) = -t (\gamma~\otimes_{br}{\mathbb I}_2)\cdot({\mathbb I}_2~\otimes_{br}\gamma) =-t (\gamma~\otimes_{br}\gamma).&
\eea

In terms of the (\ref{gparametrization}) parametrization we set $t=-e^{2i\pi g}$.\par
By introducing a suitably defined 
intertwining operator $W_t$, the above braiding relation can be expressed in terms of an ordinary tensor product $\otimes$ through the positions
\bea\label{matrixrepof}
 (\gamma~\otimes_{br}{\mathbb I}_2)\mapsto \gamma\otimes {\mathbb I}_2, &\qquad
 ({\mathbb I}_2~\otimes_{br}\gamma)\mapsto W_t\otimes \gamma.
\eea
The mappings
\bea
 ({\mathbb I}_2~\otimes_{br}\gamma)\cdot 
 (\gamma~\otimes_{br}{\mathbb I}_2)&\mapsto& (W_t\otimes\gamma)\cdot(\gamma\otimes {\mathbb I}_2)= (W_t\gamma)\otimes\gamma\nonumber\\
 (\gamma~\otimes_{br}{\mathbb I}_2)\cdot ({\mathbb I}_2~\otimes_{br}\gamma)&\mapsto&
(\gamma\otimes {\mathbb I}_2)\cdot (W_t \otimes\gamma) = (\gamma W_t)\otimes\gamma
\eea
imply that the consistency condition for the $2\times 2$ interwining operator $W_t$ is given by
\bea\label{consistencyintertwining}
W_t\gamma &=& (-t) \gamma W_t.
\eea
A solution, expressed in terms of the  $t=-e^{2i\pi g}$ position, is 
\bea\label{solutionintertwining}
W_t&=& \cos(-\pi g)\cdot {\mathbb I_2} + i \sin(-\pi g)\cdot X,\qquad {\textrm{where $X=$ {\footnotesize  $\left(\begin{array}{cc} 1&0\\0&-1\end{array}\right)$.}}}
\eea
Indeed, we get
\bea
W_t\gamma &=& e^{2\pi i g}\gamma W_t.
\eea
The new building blocks of the braided $2$-particle ($2P$) and $3$-particle ($3P$) creation operators
can be respectively defined as
\bea\label{braidedcr23}
2P &:& A_1^\dagger := \gamma\otimes {\mathbb I}_2, \quad \qquad A_2^\dagger :=W_t \otimes \gamma;  \nonumber\\
3P &:& B_1^\dagger := \gamma\otimes {\mathbb I}_2\otimes{\mathbb I}_2, \quad B_2^\dagger :=W_t \otimes \gamma\otimes{\mathbb I}_2,\quad
B_3^\dagger := W_t\otimes W_t\otimes \gamma.
\eea
They are nilpotent, due to the nilpotency of the $\gamma$ matrix:
\bea\label{newnil}
 (A_i^\dagger)^2=0 \quad{\textrm{for $i=1,2$}} \qquad {\textrm{and}}\qquad 
 (B_j^\dagger)^2=0 \quad{\textrm{for $j=1,2,3$.}}
\eea
Their respective braiding relations are
\bea\label{newbraid2p}
&(A_2^\dagger\cdot A_1^\dagger )= e^{2\pi i g}(A_1^\dagger\cdot A_2^\dagger)&
\eea
and
\bea\label{newbraid3p}
&(B_3^\dagger\cdot B_1^\dagger )= e^{2\pi i g}(B_1^\dagger\cdot B_3^\dagger), \quad (B_3^\dagger\cdot B_2^\dagger )= e^{2\pi i g}(B_2^\dagger\cdot B_3^\dagger), \quad (B_2^\dagger\cdot B_1^\dagger )= e^{2\pi i g}(B_1^\dagger\cdot B_2^\dagger).  &
\eea
The extension to the braiding of creation operators for $N>3$  particle sectors is straightforward. It requires $N-1$ tensor products and produces $2^N\times 2^N$ matrices; we can define
\bea\label{newbuildingn}
A_{k;N}^\dagger &:=&{\underbrace{W_t\otimes \cdots\otimes W_{t}}_{k-1}} \otimes \gamma\otimes {\underbrace{{\mathbb I}_2\otimes \cdots \otimes {\mathbb I}_2}_{N-k}} \qquad {\textrm{for~~ $k=1,2,\ldots, N$.}}
\eea
The $N$-particle vacuum state is expressed by the $2^N$ column vector $|vac\rangle_N$ given by
\bea\label{newnvacuum}
|vac\rangle_N ^T&=& (1,0,0,\ldots, 0), \quad {\textrm{where only the first entry is nonvanishing.}}
\eea
By construction, the new (\ref{newbuildingn}) building blocks $A_{k;N}^\dagger$ are nilpotent  ($(A_{k;N}^\dagger)^2=0$) and satisfy, for $k<k'$, the relations $A_{k';N}^\dagger A_{k;N}^\dagger= e^{2\pi i g} A_{k;N}^\dagger A_{k';N}^\dagger$.\par
~\par
The $N$-particle Hamiltonians $H_N$  coincide with the matrices defined in (\ref{nparticlehamiltonians}).\par
~\par
It should be pointed out, for later convenience, that the $A_k^\dagger, A_k$ building blocks of the $2$-particle creation and annihilation operators belong to a non-standard (i.e., non block-diagonal, see \cite{dggt1}) odd sector
of a ${\mathbb Z}_2$-graded decomposition of the $4\times 4$ matrices. In the non-standard decomposition
of the $4\times 4$ matrices the nonvanishing entries (denoted with the symbol ``$\ast$")  of the even (odd) sector $M_0$ ($M_1$) are accommodated in
{\footnotesize{\bea \label{nsmatrix}
M_0 \equiv \left(\begin{array}{cccc} \ast&0&0&\ast\\0&\ast&\ast&0\\0&\ast&\ast&0\\\ast&0&0&\ast\end{array}\right), &&
M_1 \equiv \left(\begin{array}{cccc} 0&\ast&\ast&0\\\ast&0&0&\ast\\\ast&0&0&\ast\\0&\ast&\ast&0\end{array}\right).
\eea}} 
The ${\mathbb Z}_2$ grading is respected since
\bea
M_i\cdot M_j' = M_{i+j}''&& {\textrm{for ~}} i,j=0,1, \quad {\textrm{with ~}}~ i+j= 0,1 ~~mod ~2.
\eea 
From (\ref{braidedcr23}) the $2$-particle creation building blocks $A_1^\dagger, A_2^\dagger$ and their respective conjugate matrices $A_1, A_2$  are
{\footnotesize{\bea \label{nonstandard2p}
A_1^\dagger =\left(\begin{array}{cccc} 0&0&0&0\\0&0&0&0\\1&0&0&0\\0&1&0&0\end{array}\right),  &&
A_2^\dagger= \left(\begin{array}{cccc} 0&0&0&0\\e^{-i\pi g}&0&0&0\\0&0&0&0\\0&0&e^{i\pi g}&0\end{array}\right), \nonumber\\
A_1 =\left(\begin{array}{cccc} 0&0&1&0\\0&0&0&1\\0&0&0&0\\0&0&0&0\end{array}\right),  &&
A_2= \left(\begin{array}{cccc} 0&e^{i\pi g}&0&0\\0&0&0&0\\0&0&0&e^{-i\pi g}\\0&0&0&0\end{array}\right).
\eea}} 
An even $4\times 4$ central charge $c$ is defined as
\bea \label{centralcharge}
c&=& diag(1,1,1,1).
\eea
As shown in Section {\bf 8},  the $4$ matrices in (\ref{nonstandard2p}) and the central charge $c$ close
a generalized ``mixed-bracket" Heisenberg-Lie algebra.

\section{Indistinguishability as a superselection}

 Let us consider the braided set (\ref{newbuildingn}) of $N$-particle creation operators $A_{k;N}^\dagger$ ($k=1,2,\ldots, N$); the operator $A_{k;N}^\dagger$ produces, when applied to the (\ref{newnvacuum}) vacuum state $|vac\rangle_N$,  an excited state in the $k$-th 
position. Two scenarios can be considered. In the first scenario the particles in the $k$-th and $k'$-th ($k'\neq k$) positions are assumed to be of different type;  in the alternative second scenario they are of the same type. The latter case brings the notion of quantum indistinguishability which implies, following the discussion of Section {\bf 2}, that the powers of a unique particle creation operator have to be applied. The analogous of formula (\ref{general}) for the (\ref{newbuildingn}) building blocks  constructed in terms of the interwining operator $W_t$ reads as
\bea
{A^{(N)}}^\dagger&:=&
\sum_{k=1}^NA_{k;N}^\dagger.
\eea
The truncation phenomenon discussed in Section {\bf 2} for the root-of-unity braidings is implied by the  $({A^{(N)}}^\dagger)^n$ powers of this composite operator.  The indistinguishability case can be recovered from the first scenario by applying a superselection of the observables. This feature is easily illustrated in the $2$-particle setting for the third-root-of unity. It is immediate to generalize this construction to arbitrary $N$-particle sectors and arbitrary values of the $g$ parameter defining, under the $t=-e^{2i\pi g}$ identification, $W_t$ in  formula (\ref{solutionintertwining}).\par
~\par
For the third root of unity of the $2$-particle sector we specialize the two $4\times 4$ creation operators $A_1^\dagger, A_2^\dagger$ 
from (\ref{nonstandard2p}) as given by $g=1/3$. By setting
\bea\label{specialize}
{\overline{A}}_1^\dagger = {A_1^\dagger}_{|{g=\frac{1}{3}}},&&
{\overline{A}}_2^\dagger = {A_2^\dagger}_{|{g=\frac{1}{3}}},
\eea
we obtain
{\footnotesize{\bea \label{specialize2}
{\overline A}_1^\dagger =\left(\begin{array}{cccc} 0&0&0&0\\0&0&0&0\\1&0&0&0\\0&1&0&0\end{array}\right),  &&
{\overline A}_2^\dagger= \left(\begin{array}{cccc} 0&0&0&0\\e^{-i\pi /3}&0&0&0\\0&0&0&0\\0&0&e^{i\pi /3}&0\end{array}\right).
\eea}} 
These creation operators satisfy
\bea
{\overline A}_2^\dagger\cdot{\overline  A}_1^\dagger= j{\overline  A}_1^\dagger \cdot{\overline  A}_2^\dagger \qquad {\textrm{for $j=e^{i\frac{2}{3}\pi}$.}}
\eea
The four basis vectors spanning the Hilbert space ${\cal H}_{dist}$ of $2$ distinct particles can be given by
\bea
&v_{00}, \quad v_{10} =  {\overline A}_1^\dagger v_{00}, \quad v_{01} = {\overline A}_2^\dagger v_{00}, \quad v_{11} = 
{\overline A}_1^\dagger {\overline A}_2^\dagger v_{00}, \quad {\textrm{ where ~ $v_{00}^T = (1,0,0,0)$.}} &\nonumber\\&&
\eea
The two-particle Hamiltonian is $H_{2}=diag(0,1,1,2)$.\par
The associated Hilbert space of indistinguishable particles is the $3$-dimensional Hilbert subspace ${\cal H}_{ind}\subset {\cal H}_{dist}$ spanned by the following $v_{E}$ energy eigenvectors with respective eigenvalues $E=0,1,2$:
\bea\label{indist1}
&v_{E=0} = v_{00},\qquad v_{E=1} = {\overline{A}}^\dagger v_{00}, \qquad v_{E=2}= ({\overline{A}}^{\dagger})^2 v_{00},\qquad 
{\textrm{for ~~ ${\overline A}^\dagger = {\overline A}_1^\dagger +{\overline A}_2^\dagger$, }}
&
\eea
so that
\bea\label{indist2}
&v_{E=0} = v_{00},\qquad v_{E=1} = v_{10}+v_{01},\qquad v_{E=2}= -j^2v_{11}.
&
\eea
The indistinguishable Hilbert space ${\cal H}_{ind}$ is recovered from the superselection induced by the Hermitian projector operator ${\overline P}$ given by
\bea
&{\overline P}= \left(\begin{array}{cccc} 1&0&0&0\\0&\frac{1}{2}&-\frac{1}{2}j&0\\0&-\frac{1}{2}j^2&\frac{1}{2}&0\\0&0&0&1\end{array}\right), \qquad {\textrm{where ~~${\overline P}^2={\overline P}$ ~~ and ~~ ${\overline P}^\dagger = {\overline P}$.}}&
\eea
The projector ${\overline P}$ commutes with the Hamiltonian ${H}_2$:
\bea
\relax [ {\overline P} , {H}_{2}] &=&0.
\eea
The superselected Hilbert subspace spanned by the $+1$ eigenvectors of ${\overline P}$
coincides with the Hilbert space ${\cal H}_{ind}$. Indeed, we get
\bea\label{superselected}
{\overline P} \cdot v_{E=0,1,2} &=& v_{E=0,1,2}, \qquad \quad {\textrm{while ~~ ${\overline P}\cdot (v_{10}-v_{01} )=0$.}}
\eea

\section{About Volichenko algebras and metasymmetries}

The notion of ``symmetries wider than supersymmetry" was stressed in the Leites-Serganova \cite{lese1} paper and their further \cite{lese2} work. A key point that they mentioned is the possibility of statistics-changing maps  which
do not preserve the grading of ordinary superalgebras.\par
As an implementation of this scheme they introduced the notion of {\it Volichenko algebras} (after a theorem proved by Volichenko) as nonhomogeneous subspaces of Lie superalgebras closed under the superbracket. In \cite{lese1} the possible application to parastatistics was pointed out;  the concept of superspace was enlarged to {\it metaspace} to take into account the class of transformations which do not preserve the grading. Similarly, the concept of symmetry or supersymmetry was enlarged to that of ``metasymmetry". \par
Technically, a Volichenko algebra satisfies a condition which is known as ``metaabelianess"; this means that any triple of $x,y,z$ generators satisfies the identity
\bea\label{metabel}
\relax [x,[y,z]] &=& 0
\eea
where, in the above relations, the ordinary commutators are taken.\par
An intriguing example of a class of Volichenko algebras was presented in \cite{lese2}. Denoted as ${\mathfrak{vgl}}_\mu(p|q)$, they are given by the ${\mathfrak{gl(p|q)}}$ space of $(p+q)\times(p+q)$ supermatrices. Under the ${\mathbb Z}_2$-decomposition ${\mathfrak{gl(p|q)}}={\mathfrak h}_0\oplus {\mathfrak h}_1$ into even/odd sectors, by fixing $\mu=a/b\in {\mathbb C}P^1$, a multiplication ${\mathfrak h}_1\times {\mathfrak h}_1\rightarrow {\mathfrak h}_0$ is introduced through the formula
\bea
(x,y)_\mu&=& a[x,y]+b\{x,y\}\qquad {\textrm{for any $x,y\in {\mathfrak{h}}_1$.}}
\eea
For $ab\neq 0$ the $(.,.)_\mu$ bracket is an interpolation of ordinary commutators and anticommutators.\par
~
\par
This particular example was my starting point to explore the possibility of introducing  ``mixed brackets" to describe the parastatistics of the braided Majorana qubits. There were two main reasons for that:\\
~\\
{\it i}) a single-particle Majorana qubit is created/annihilated by the $\gamma,\beta$ generators in the odd sector of the ${\mathfrak{gl}}(1|1)$ superalgebra, as shown in formulas (\ref{4op}, \ref{anticomm});\\
{\it ii}) the building blocks of the $2$-particle creation/annihilation operators are accommodated in the odd sector
of a ${\mathfrak{gl}}(2|2)$ superalgebra, see formula (\ref{nonstandard2p}), under the (\ref{nsmatrix}) non-standard decomposition into even/odd matrices. \\
~\\
These considerations lead to the introduction of ``mixed-bracket" generalized Heisenberg-Lie algebras which close 
generalized Jacobi identities (such new algebras are presented in Section {\bf 8}).
To my surprise, these algebras do not satisfy, as it can be easily checked, the metaabelianess condition (\ref{metabel})
for any triple of generators; therefore, they are not Volichenko algebras according to the Leites-Serganova's definition.
Despite of that, they present a closed mixed-bracket structure; furthermore, as shown in Section {\bf 10},  the generators of the generalized Heisenberg-Lie algebras define a mixed-bracket closed dynamical symmetry of a matrix Partial Differential Equation.\par
~\\
This leads to a relevant consideration which will be further discussed in the Conclusions.  It points out towards a relaxation of the Leites-Serganova's notion of ``metasymmetry" which allows to accommodate mixed-bracket structures that do not
necessarily satisfy the Volichenko metaabelianess condition. A quote from the original \cite{lese1} paper seems to provide the solution: ``{\it This feature of Volichenko algebras could be significant for parastatistics because once we abandon the Bose-Fermi statistics there seem to be too many ad hoc ways to generalize}". \\
As a matter of fact, the mathematical structure of a simple model like the one here investigated (the multi-particle braided Majorana qubits) implies a parastatistics which has all key ingredients: mixed brackets, transformations and symmetries which do not preserve the ${\mathbb Z}_2$-grading of superalgebras, etc.; for this model it is the dynamics of the braided multi-particle sector  which forces the introduction of a parastatistics  which does not fulfill the metaabelianess condition.  This means that the mathematical structure is not obtained from an ad hoc prescription, but it is implied by the multi-particle dynamics of the model. It is shown in Section {\bf 8} that
the mixed-bracket Heisenberg-Lie algebras satisfy a new identity which can be regarded as a metaabelianess condition expressed in terms of the mixed brackets.\par
~\par
Concerning Volichenko algebras, historical motivations and references leading to their introduction are found in \cite{{lese1},{lese2}}. A construction of Volichenko algebras as algebras of differential operators is found in
\cite{iye}. A recent account with updated references on Volichenko algebras is \cite{lei}.

\section{Braided qubits and mixed-bracket Heisenberg-Lie algebras}

As mentioned in the previous Section, the Leites-Serganova notion of ``symmetries wider than supersymmetry" applies to statistics-changing maps; it is the source of inspiration for introducing  a mixed-bracket structure for
the multi-particle creation/annihilation operators of the braided Majorana qubits. The construction, which is here presented, implies a mixed bracket generalization of the Heisenberg-Lie algebras.\par
Let $X,Y$ be two operators. Their mixed-bracket, defined in terms of a $\vartheta_{XY}$ angle and denoted as $(X,Y)_{\vartheta_{XY}}$, is an interpolation of the ordinary $[X,Y]$ commutator and  $\{X,Y\}$ anticommutator. We can set
\bea\label{volimixedbracket}
(X,Y)_{\vartheta_{XY}} &:=& i \sin(\vartheta_{XY}) \cdot [X,Y] + \cos(\vartheta_{XY})\cdot \{X,Y\},
\eea
where $\vartheta_{XY}$ belongs, $mod~   2\pi$, to the interval $\vartheta_{XY}\in [-\pi,\pi[$.\\
The following identity holds:
\bea
&(Y,X)_{\vartheta_{YX}}= (X,Y)_{\vartheta_{XY}}\qquad {\textrm{for ~~~$\vartheta_{YX}=-\vartheta_{XY}$}}.&
\eea
With the above definitions, if $X,Y$ are Hermitian operators, the right hand side is also Hermitian:
\bea
X^\dagger=X,\quad  Y^\dagger = Y&\rightarrow & (X,Y)_{\vartheta_{XY}}^\dagger = (Y,X)_{\vartheta_{YX}} =(X,Y)_{\vartheta_{XY}} .
\eea
We are now in the position to introduce the generalized, mixed-bracket, Heisenberg-Lie algebras of level-$s$ for the 
$2$-particle and $3$-particle sectors. 
Due to the coassociativity of the (\ref{coproductbr}) coproduct, the extension to the induced  mixed-bracket generalized Heisenberg-Lie algebras for any $N$-particle sector is immediate.\par
~\\
{\it Remark 1}:  the $2$-particle level$-s$ generalized Heisenberg-Lie algebra consists of the $5$ generators entering formulas
(\ref{nonstandard2p}) and (\ref{centralcharge}) with the position $g=\frac{1}{s}$. The $5$ generators are the $4\times 4$ matrices $A_1,A_2, A_1^\dagger, A_2^\dagger$ and the central charge $c=diag(1,1,1,1)$.\par
~\\
{\it Remark 2}: the $3$-particle level$-s$ generalized Heisenberg-Lie algebra admits $7$ generators which are presented as $8\times 8$ matrices.  They are the three creation operators $B_1^\dagger, B_2^\dagger, B_3^\dagger$ introduced, 
with the position $g=\frac{1}{s}$, in 
the second line of formula (\ref{braidedcr23}),  together with their respective adjoint operators $B_1,B_2,B_3$ ($B_i = (B_i^\dagger)^\dagger$ for $i=1,2,3$) and the central charge $c_3=diag(1,1,1,1,1,1,1,1)$.\par
~\par
At a formal level, the $2$-particle and $3$-particle mixed-bracket Heisenberg-Lie algebras can be recovered from, respectively, $2$-oscillator and $3$-oscillator Heisenberg-Lie algebras by replacing the ordinary (anti)commutators with the mixed brackets, so that  $[.,.] \rightarrow (.,.)$ for appropriate choices of the angles.
\par
~\par
Let us now specify the angles and introduce the closed algebraic structures.\par
~\par
It is convenient, for the $2$-particle sector, to rename the generators as
\bea
&G_0:=c=diag(1,1,1,1),\quad G_{+1}:=A_1^\dagger,\quad G_{-1}:= A_1, \quad G_{+2} := A_2^\dagger, \quad G_{-2}:= A_2. &
\eea
In principle a suffix $s$ should be introduced to specify the level of the truncation; it will be omitted in order not to burden the notation.\par
The Hermiticity conditions are
\bea
&G_0^\dagger = G_0, \qquad  G_{\pm 1}^\dagger = G_{\mp 1}, \qquad G_{\pm 2}^\dagger = G_{\mp 2}.&
\eea

The  ${\mathbb Z}_2$-grading (\ref{nsmatrix}) of the $4\times 4$ matrices implies that the generator $G_0$ is even ($\varepsilon=0$), while the four generators $G_{\pm1}, G_{\pm 2}$ are odd ($\varepsilon=1$).\par
~\par
For $I,J =0,\pm1, \pm2$, a suitable choice of the $\vartheta_{IJ}$  angles giving a closed structure to the
$ 
(G_I,G_J)_{\vartheta_{IJ}} = i \sin(\vartheta_{IJ}) \cdot [G_I,G_J] + \cos(\vartheta_{IJ})\cdot \{G_I,G_J\} $
brackets is the following.\par
~\par
The vanishing brackets involving the central element $G_0$ coincide with an ordinary commutator; for any $I=0,\pm1, \pm2 $
we have
\bea\label{g0bracket}
(G_0, G_I)_{\pm \frac{\pi}{2}} =  
(G_I, G_0)_{\mp \frac{\pi}{2}} =0 &\Leftrightarrow  & [G_0,G_I]=[G_I,G_0]=0.
\eea
~\par
The extra angles entering the $(G_I,G_J)_{\vartheta_{IJ}}$ brackets for $I,J=\pm 1 ,\pm 2$ are determined by the following construction.
Let us introduce the $N^L, N^R$ diagonal operators
\bea
N^L =-\frac{1}{2}\cdot diag(1,1,-1,-1), &&  
N^R =-\frac{1}{2}\cdot diag(1,-1,1,-1)
\eea
such that 
\bea
[N^L, G_I] =\lambda_I^L G_I, &&
[N^R, G_I] =\lambda_I^R G_I,
\eea
where
\bea
\lambda^L_{\pm 1} =\pm1, \quad 
\lambda^L_{\pm 2} =0, &\qquad&
\lambda^R_{\pm 1} =0, \quad 
\lambda^R_{\pm 2} =\pm 1.
\eea
The further parameters $\mu_I, \nu_I$ are introduced through the positions
\bea
\mu_I = \lambda^L_I+\lambda^R_I, && \nu_I=(\lambda^L_I)^2-(\lambda^R_I)^2.
\eea
Their corresponding values for the $G_I$ generators are read from the table
\bea
&\begin{array}{|c|c|c|} \hline &\mu_I&\nu_I\\ 
\hline 
G_{+1}:&+1&+1\\ \hline 
G_{-1}:&-1&+1\\ \hline
G_{+2}:&+1&-1\\ \hline
G_{-2}:&-1&-1\\ \hline\end{array}&
\eea
~\par
At the level-$s$ ($s=2,3,4,5,\ldots$), the angles $\vartheta_{IJ}$ for $I,J=\pm 1,\pm 2$ are determined by the formula
\bea
\vartheta_{IJ} = k_s \cdot\mu_I\mu_J\cdot (\nu_I-\nu_J) &\qquad& {\textrm{for ~~ $k_s = \frac{s+2}{4s}\pi$.}}
\eea
As mentioned before, this construction formally resembles a $2$-oscillator algebra with the only nonvanishing brackets given by
\bea
&(G_{\pm 1}, G_{\mp 1}) =  (G_{\pm 2}, G_{\mp 2}) = G_0.&
\eea
The mixed-bracket formulas, with the  explicit insertion of the $\vartheta_{IJ}$ angle dependence, are
\bea\label{anticommut}
&(G_{+1},G_{-1})_0=(G_{-1},G_{+1})_0= G_0, \qquad (G_{+1},G_{+1})_0=(G_{-1},G_{-1})_0 =0,&\nonumber\\
&(G_{+2},G_{-2})_0=(G_{-2},G_{+2})_0= G_0, \qquad (G_{+2},G_{+2})_0=(G_{-2},G_{-2})_0=0,&
\eea
together with 
\bea\label{genuinemixed}
&(G_{+1},G_{+2})_{+\frac{s+2}{2s}\pi}=(G_{+2},G_{+1})_{-\frac{s+2}{2s}\pi}= 0, &\nonumber\\
&(G_{+1},G_{-2})_{-\frac{s+2}{2s}\pi}=(G_{-2},G_{+1})_{+\frac{s+2}{2s}\pi}= 0, &\nonumber\\
&(G_{-1},G_{+2})_{-\frac{s+2}{2s}\pi}=(G_{+2},G_{-1})_{+\frac{s+2}{2s}\pi}= 0, &\nonumber\\
&(G_{-1},G_{-2})_{+\frac{s+2}{2s}\pi}=(G_{-2},G_{-1})_{-\frac{s+2}{2s}\pi}= 0. &
\eea
~\par
The brackets entering (\ref{anticommut}) are ordinary anticommutators, while the brackets expressed by
(\ref{genuinemixed}) are, for $s>2$,  a linear combination of commutators and anticommutators.\par
~\par
At $s=2$ the (\ref{g0bracket},\ref{anticommut},\ref{genuinemixed}) brackets define an ordinary, $2$ fermionic oscillators, Heisenberg-Lie algebra.\par
For any given $s=3,4,5,\ldots$, the (\ref{g0bracket},\ref{anticommut},\ref{genuinemixed}) brackets define
a closed, $5$-generator, mixed-bracket algebra which will be denotes as the ``$2$-particle, mixed-bracket Heisenberg-Lie algebra (of level-$s$)".\par
~\par
Since $G_0$ is a central element, these level$-s$ mixed-bracket algebras not only satisfy generalized Jacobi identities; they satisfy the
stronger identity
\bea
(G_I,(G_J,G_K)) &=& 0 \qquad {\textrm{for any ~~$I,J,K=0,\pm 1,\pm 2$
}}.
\eea
 By expressing the angle dependence, the above identity reads as
\bea\label{newmeta}
(G_I,(G_J,G_K)_{\vartheta_{JK}})_{\vartheta_{I,J+K}} &=& 0,
\eea
where $G_0$, the only possible nonvanishing element for $(G_J,G_K)$, is recovered from $J+K=0$. \par
~\par
One should compare the similarity of the expression (\ref{newmeta}) with the metaabelianess condition (\ref{metabel}).  The (\ref{metabel}) condition involves only ordinary commutators, while the (\ref{newmeta}) identity
involves the $(.,.)$ mixed brackets. It is quite tempting to relax, as discussed in Section {\bf 7}, the  metaabelianess condition presented in \cite{lese1} with a new metaabelianess criterium based  on the identity (\ref{newmeta}) which involves mixed brackets.\par
~\par
One can indeed easily check that, while the set of $5$ generators $G_I$ for $I=0,\pm 1,\pm2$ always satisfy (\ref{newmeta}),  some combinations of the operators violate the (\ref{metabel}) metaabelianess condition.  A violation is given, e.g., by the expression
\bea
\relax [G_{+1},[G_{+2},G_{-2}]] = 2i \sin(\frac{\pi}{s}) \left(E_{21}-E_{43}\right) &\neq &0,
\eea
where $E_{ij}$ denotes the $4\times 4$ matrix with entry $1$ at the intersection of the $i^{th}$ row with the $j^{th}$
column and $0$ otherwise. \par
~\par
We present, for completeness, the $3$-particle sector of the level-$s$ mixed-bracket Heisenberg-Lie algebras.
As done for the $2$-particle case, it is convenient to rename the $7$ generators which were previously introduced in the above {\it Remark 2}. Let us set 
\bea
&{\overline G}_0:=c_3=diag(1,1,1,1,1,1,1,1),\quad {\overline G}_{+i}:=B_i^\dagger,\quad {\overline G}_{-i}:= B_i, \qquad {\textrm{for~~$i=1,2,3$.}} &
\eea
The brackets defined for ${\overline G}_0$ are ordinary commutators:
\bea\label{g0bracket3p}
({\overline G}_0, {\overline G}_I)_{\pm \frac{\pi}{2}} =  
({\overline G}_I, {\overline G}_0)_{\mp \frac{\pi}{2}} =0,  \qquad {\textrm{for ~~ $I=0,\pm1,\pm 2,\pm3$.}} &
\eea
The following brackets, defined for $i=1,2,3$, are ordinary anticommutators:
\bea
\label{anticomm3}
&({\overline G}_{+i},{\overline G}_{-i})_0=({\overline G}_{-i},{\overline G}_{+i})_0= {\overline G}_0, \qquad ({\overline G}_{+i},{\overline G}_{+i})_0=({\overline G}_{-i},{\overline G}_{-i})_0 =0.&
\eea
There are three sets ($S_{12}, S_{13}, S_{23}$) of mixed brackets, respectively involving the generators:
\bea
&S_{12}:~ {\overline G}_{\pm 1} ~~{\textrm{and ~}} {\overline G}_{\pm 2}; \qquad   S_{13}: ~{\overline G}_{\pm 1} ~~{\textrm{and ~}} {\overline G}_{\pm 3};\qquad S_{23}:~ {\overline G}_{\pm 2} ~~{\textrm{and ~}} {\overline G}_{\pm 3}.  &
\eea
The respective mixed brackets are recovered, {\it with the same values of the angles}, from fomula (\ref{genuinemixed})
under the positions
\bea\label{mixed3}
S_{12}&:&  G_{\pm 1}\mapsto  {\overline G}_{\pm 1},\quad G_{\pm 2}\mapsto  {\overline G}_{\pm 2},\nonumber\\
S_{13}&:&  G_{\pm 1}\mapsto  {\overline G}_{\pm 1},\quad G_{\pm 2}\mapsto  {\overline G}_{\pm 3},\nonumber\\
S_{23}&:&  G_{\pm 1}\mapsto  {\overline G}_{\pm 2},\quad G_{\pm 2}\mapsto  {\overline G}_{\pm 3}.
\eea
The three $5$-generator subalgebras, respectively spanned by the sets of generators
$\{ {\overline G}_0, {\overline G}_{\pm1}, {\overline G}_{\pm 2}\}$, $\{ {\overline G}_0, {\overline G}_{\pm1}, {\overline G}_{\pm 3}\}$ and $\{ {\overline G}_0, {\overline G}_{\pm 2}, {\overline G}_{\pm 3}\}$, are all isomorphic to the level-$s$, $2$-particle, mixed-bracket Heisenberg-Lie algebras.\par
~\par
As mentioned before, the extension of the construction to $N$-particle mixed-bracket Heisenberg-Lie algebras for any integer $N>3$ is immediate.
\par

\section{Color parafermionic oscillators from the $s\rightarrow \infty$ limit}

The level-$s$ mixed-bracket Heisenberg-Lie algebras provide an interpolation between the ordinary $s=2$ fermionic oscillators and the untruncated spectrum case recovered in the $s\rightarrow\infty $ limit. Since the $s=\infty$ spectrum is untruncated one can loosely refer, with a caveat, to this case as ``bosonic". The caveat is that in both truncated and untruncated cases the original ${\mathbb Z}_2$-grading holds; the creation operators are odd and the spectrum 
is split into even (including the vacuum) and odd states. The superselection rule which holds for ordinary bosons/fermions is still in place: therefore one can only makes superpositions (linear combinations) either of even states  
or of odd states (without mixing them). The most general state in the ${\mathbb Z}_2$-graded Hilbert space is either $\varphi=\sum_a c_a\varphi_a$ (the sum being taken on the orthonormal basis of the even states) or $\psi=\sum_\alpha c_\alpha\psi_\alpha$ (the sum being taken on the orthonormal basis of the odd states). This point was discussed at length in \cite{topqubits}.\par
For a single Majorana qubit this implies that the ${\mathbb Z}_2$-graded Hilbert space is ${\mathbb C}^{1|1}$, instead of ${\mathbb C}^2$ of an ordinary qubit. The analogous of the Bloch sphere for a Majorana qubit is the ``${\mathbb Z}_2$-sphere" expressed by two points respectively identified with $+1$ or $-1$. The space of the physically inequivalent configurations individuated by the ray vectors corresponds to $1$ bit of information. We have an even state (labeled as ``$0$" in a suitable parametrization) associated with the vacuum ray vector and an odd state (labeled as ``$1$") associated with the excited Majorana ray vector.\par
On the basis of the (\ref{genericenergy}) result presented in Section {\bf 3}, the ${\mathbb Z}_2$-graded Hilbert spaces of the $N$-particle sectors are given, in the $s\rightarrow \infty$ untruncated limit, by
\bea\label{gradedspaces}
{\mathbb C}^{n+1|n+1},&&{\textrm{for an odd integer~ $N= 2n +1$  ~ and}}\nonumber\\
{\mathbb C}^{n+1|n},&&{\textrm{for an even integer $N= 2n$.}}
\eea

In the $s\rightarrow \infty$ limit the brackets recovered from formulas 
(\ref{g0bracket},\ref{anticommut},\ref{genuinemixed}) in the $2$-particle sector
and  from formulas (\ref{g0bracket3p},\ref{anticomm3},\ref{mixed3}) in the $3$-particle sector are ordinary commutators and anticommutators (that is, there are no mixed terms since either the $\sin$ function or the $\cos$ function entering the right hand side of (\ref{volimixedbracket}) is vanishing). Despite being ordinary commutators/anticommutators,
the brackets are {\it arranged in a peculiar way}. To illustrate this point we present the $2$-particle case.\par
~\par
In the $s\rightarrow \infty$ limit the five $4\times 4$ matrices $G_I$ ($I=0,\pm1 ,\pm2$) are given by
{\footnotesize{\bea \label{slimit}
G_0=\left(\begin{array}{cccc} 1&0&0&0\\0&1&0&0\\0&0&1&0\\0&0&0&1\end{array}\right),\qquad  G_{+1} =\left(\begin{array}{cccc} 0&0&0&0\\0&0&0&0\\1&0&0&0\\0&1&0&0\end{array}\right),  &&
G_{+2}= \left(\begin{array}{cccc} 0&0&0&0\\1&0&0&0\\0&0&0&0\\0&0&1&0\end{array}\right), \nonumber\\
G_{-1} =\left(\begin{array}{cccc} 0&0&1&0\\0&0&0&1\\0&0&0&0\\0&0&0&0\end{array}\right),  &&
G_{-2}= \left(\begin{array}{cccc} 0&1&0&0\\0&0&0&0\\0&0&0&1\\0&0&0&0\end{array}\right).
\eea}} 
Both sets of $\{G_0,G_{\pm 1}\}$ and $\{G_0,G_{\pm 2}\}$ subalgebras close a fermionic oscillator algebra:
\bea\label{paraf1}
\relax &[G_0, G_{\pm 1}] =0, \qquad \{G_{\pm 1}, G_{\pm 1} \}=0, \qquad \{G_{+1}, G_{-1}\} = G_0,&\nonumber\\
\relax &[G_0, G_{\pm 2}] =0, \qquad \{G_{\pm 2}, G_{\pm 2} \}=0, \qquad \{G_{+2}, G_{-2}\} = G_0.&
\eea
Unlike the $s=2$ case, whose full algebra corresponds to $2$ fermionic oscillators, in the $s\rightarrow \infty $ limit
the brackets taken for one $G_{\pm 1}$ and one $G_{\pm 2}$ generator are given by ordinary commutators:
\bea\label{paraf2}
\relax [G_{\pm 1}, G_{\pm 2}] =0, && [G_{\pm 1}, G_{\mp 2}]=0.
\eea
The (\ref{paraf1},\ref{paraf2}) brackets define the ${\mathbb Z}_2^2$-graded parafermionic oscillators algebra.\par~\par

Let's see in more detail the ${\mathbb Z}_2^2$-graded construction. ${\mathbb Z}_2^2$-graded ``color" Lie algebras and superalgebras were introduced by Rittenberg-Wyler in \cite{{riwy1},{riwy2}} (see also \cite{sch}) as extensions of ordinary Lie algebras and superalgebras. The generators of a color (super)algebra ${\mathfrak g}$ are accommodated in a ${\mathbb Z}_2^2$ grading:
\bea
{\mathfrak g}&=&{\mathfrak g}_{00}\oplus{\mathfrak g}_{10}\oplus{\mathfrak g}_{01}\oplus{\mathfrak g}_{11}.
\eea
A $[.,.\}$ bracket which respects the grading is introduced:
\bea
\relax &[.,.\}: {\mathfrak{g}}\times{\mathfrak{g}}\rightarrow {\mathfrak{g}}\qquad 
{\textrm{is such that \quad
$[{\mathfrak{g}}_{ij},{\mathfrak{g}}_{kl} \} \subset {\mathfrak{g}}_{i+k,j+l}$ \quad (mod $2$). }}
&
\eea
 The $[.,\}$ bracket is either a commutator or an anticommutator (one can therefore express $[{\mathfrak{g}}_{ij},{\mathfrak{g}}_{kl} \}\equiv
{\mathfrak{g}}_{ij}\cdot {\mathfrak{g}}_{kl} \pm {\mathfrak{g}}_{kl}\cdot{\mathfrak{g}}_{ij}$). The color (super)algebra satisfies a ${\mathbb Z}_2^2$-graded Jacobi identity (see  \cite{{riwy1},{riwy2}} for details).  The two consistent assignments of commutators/anticommutators for the $[.,\}$ bracket respectively define a ${\mathbb Z}_2^2$-graded color Lie algebra or a ${\mathbb Z}_2^2$-graded color Lie superalgebra. The case under consideration here is that of the color superalgebra. Its (anti)commutators assignment can be read from the entries of the table:
\bea\label{z2z2parastat}
& \begin{array}{|c||c|c|c|c|}\hline  &00&10&01&11\\ \hline \hline 00&0&0&0&0\\ \hline 10&0&1&0&1\\ \hline 01&0&0&1&1\\ \hline 11&0&1&1&0 \\ \hline \end{array}&
\eea
The entry $0$ denotes a commutator ($0\equiv [.,.]$); the entry $1$ denotes an anticommutator ($1\equiv \{.,.\}$).
One could note that the generators of the $10$-sector anticommute among themselves; the same is true  for the generators belonging to the $01$-sector. Particles associated with these sectors satisfy the Pauli exclusion principle and are therefore fermionic-like in nature. On the other hand, unlike ordinary fermions, $10$-particles commute with $01$-particles. This means that they define two classes of parafermions. Similarly, the bosonic-like particles belonging to the $11$-sector {\it anticommute} with the $10$- and $01$-parafermions: the particles of the $11$-sector are parabosons.\par
The $5$-generator closed algebraic structure defined by the (\ref{paraf1},\ref{paraf2}) brackets is a ${\mathbb Z}_2^2$-graded color Lie superalgebra. It will be denoted as ${\mathfrak{h}}_{\mathfrak{pf}}(2)$, standing for ``$2$ parafermionic oscillators Heisenberg-Lie algebra". The assignments of its generators into the ${\mathbb Z}_2^2$-graded sectors are
\bea
&G_0\in {\mathfrak{h}}_{\mathfrak{pf}}(2)_{00},  \qquad G_{\pm 1}\in {\mathfrak{h}}_{\mathfrak{pf}}(2)_{10},  \qquad G_{\pm 2}\in {\mathfrak{h}}_{\mathfrak{pf}}(2)_{01},  \qquad \emptyset \in {\mathfrak{h}}_{\mathfrak{pf}}(2)_{11},&
\eea
the color Lie superalgebra $11$-graded sector being empty.   \par
The composite generator of the Universal Enveloping Algebra given by
\bea
G_{+3} &:=& G_{+1}\cdot G_{+2}= {\footnotesize{\left(\begin{array}{cccc} 0&0&0&0\\0&0&0&0\\0&0&0&0\\1&0&0&0\end{array}\right)}}
\eea
belongs to the $11$-graded sector.\par
A ${\mathbb Z}_2^2$ grading of the $4\times 4$ matrices is introduced as follows (the possible nonvanishing entries in each sector are denoted with the symbol ``$\ast$"):
{\footnotesize{\bea \label{z2z2matrix}
&M_{00} \equiv \left(\begin{array}{cccc} \ast&0&0&0\\0&\ast&0&0\\0&0&\ast&0\\0&0&0&\ast\end{array}\right), ~~
M_{10} \equiv \left(\begin{array}{cccc} 0&0&\ast&0\\0&0&0&\ast\\\ast&0&0&0\\0&\ast&0&0\end{array}\right),~~
M_{01} \equiv \left(\begin{array}{cccc} 0&\ast&0&0\\\ast&0&0&0\\0&0&0&\ast\\0&0&\ast&0\end{array}\right),~~
M_{11} \equiv \left(\begin{array}{cccc} 0&0&0&\ast\\0&0&\ast&0\\0&\ast&0&0\\\ast&0&0&0\end{array}\right).&\nonumber\\&&
\eea}} 

{\it A comment is in order}: the above ${\mathbb Z}_2^2$ grading of the matrices implies that the Hilbert space recovered from the action of 
the $ G_{+1}, G_{+2}, G_{+3}$ matrices on the $2$-particle vacuum state $v_{00} = (1,0,0,0)^T$ can be endowed with
a ${\mathbb C}^{1|1|1|1}$ grading. The superselected indistinguishable Hilbert space, as described in Section {\bf 6}, is not compatible with the second ${\mathbb Z}_2$-graded structure, but only with the original ${\mathbb Z}_2$ grading expressed in formula (\ref{nsmatrix}). As a consequence, the $2$-particle indistinguishable Hilbert space
is three-dimensional and given, in accordance with the second formula of (\ref{gradedspaces}) for $n=1$, by ${\mathbb C}^{2|1}$. \par
~\par
Extra gradings can be introduced not only in the $s\rightarrow\infty$ limit, but also in the truncated cases. An example is presented in Appendix {\bf B} for $s=3$ (the third root of unity truncation). The introduction of a ${\mathbb Z}_3$ grading is made possible by realizing the bradings of the creation/annihilation operators via a nonminimal matrix representation.\par
~\par
In the $s\rightarrow \infty $ limit the construction of the corresponding parafermionic oscillators algebras, for any other $N$-particle sector with $N\geq 3$, is immediate.

\subsection{The $s\rightarrow \infty$ limit versus the bosonic multi-particle sectors}

It is worth discussing the difference between the $s\rightarrow \infty$ limit versus the bosonic construction of the
multi-particle sectors of the quantized model.\par
At first one should recall that the $2\times 2$ matrices $\alpha,\beta,\gamma,\delta$  introduced in (\ref{4op}) close the
${\mathfrak{gl}}(1|1)$ superalgebra with $\beta,\gamma$ odd (fermionic) elements.\par
In an alternative description the three matrices $\beta,\gamma,\alpha-\delta$ can also be regarded as the generators of the ${\mathfrak{sl}}(2)$ algebra in the fundamental representation. This description induces a bosonic quantization of the model. 

The  $N=2,3,4,\ldots$ multi-particle sectors can be constructed for both bosonic and level-$s$ fermionic versions.
In all cases the corresponding multi-particle Hilbert spaces ${\cal H}_N$  are subspaces of  ${\cal H}^{\otimes N}$, the tensor product of the two-dimensional, single-particle Hilbert space ${\cal H}$; one gets, see (\ref{subspace}), ${\cal H}_N\subset {\cal H}^{\otimes N}$. The dimension of ${\cal H}^{\otimes N}$ is $2^N$.

By taking into account that the finite-dimensional representation theory of the noncompact group $SL(2, {\mathbb R})$ is equivalent to the representation theory of $ SU(2)$,
the bosonic  formulation gives a nice interpretation of ${\cal H}^{\otimes N}$ as a decomposition into irreducible representations of the spin-$\frac{1}{2}$ reps tensors:

\bea \label{spinirreps}
&&{\cal H}^{\otimes N=2}\equiv [1/2]\otimes [1/2] = [0]\oplus[1]\Rightarrow {\textrm{dim }} 4=1+3, \nonumber\\
&&{\cal H}^{\otimes N=3}\equiv [1/2]\otimes [1/2]\otimes [1/2] = [1/2]\oplus([1/2]\oplus[3/2])\Rightarrow {\textrm{dim }} 8=2\times 2 + 1\times 4\nonumber\\
&&  {\textrm{and so on}}.
\eea

The dimensions of the fermionic (at level-$s$) and of the bosonic multi-particle Hilbert spaces  can be read from the coproducts applied to the creation operator $\gamma$.

For ordinary $(s=2)$ fermions the (\ref{tchoices}) $t=1$ choice implements,  as discussed in Section {\bf 3}, the Pauli exclusion principle: only two states (a vacuum state and an excited fermionic state) survive, for any given $N$,
in the  multi-particle sector. Therefore, for ordinary $s=2$ fermions the dimensions of the multi-particle Hilbert spaces are
\bea
&{\textrm{dim}} ({\cal H}_{N=1}) ={\textrm{dim}} ({\cal H}_{N=2})= {\textrm{dim}} ({\cal H}_{N=3})= \ldots = 2.&
\eea

Unlike the level-$s$ fermionic versions which use the $t$-dependent braided tensor products
(\ref{braidingamma}), the bosonic multi-particle quantization induced by ${\cal U }({\mathfrak {sl}}(2))$   is obtained from an ordinary (not braided) tensor product. In particular, the following  relation is satisfied:  
\bea
&({\mathbb I}_2\otimes \gamma) \cdot (\gamma\otimes {\mathbb I}_2) = (\gamma\otimes {\mathbb I}_2)\cdot ({\mathbb I}_2\otimes \gamma).&
\eea

A straightforward combinatorics shows that the dimensions of the multi-particle bosonic Hilbert spaces ${\cal H}_{bos; N}\subset {\cal H}^{\otimes N}$ are given by
\bea
&{\textrm{dim}} ({\cal H}_{bos;N}) =N+1.&
\eea
The bosonic coproduct selects, within ${\cal H}^{\otimes N}$, the highest spin subspace in the (\ref{spinirreps}) irreducible decompositions. One gets
\bea
\relax &{\textrm{dim}} ({\cal H}_{bos;N=2}) \equiv [1], \quad  {\textrm{dim}} ({\cal H}_{bos;N=3}) \equiv [\frac{3}{2}]\quad {\textrm{and, in general,}} \quad {\textrm{dim}} ({\cal H}_{bos;N}) \equiv [\frac{N}{2}].&\nonumber\\&&
\eea

It is important to notice that, for any $N$, the dimension of the bosonic $N$-particle Hilbert space coincides with the dimension of the fermionic untruncated $N$-particle Hilbert space recovered in the $s\rightarrow \infty $ limit. This feature, on the other hand, does not imply that these two $N$-particle Hilbert spaces should be identified. \par

The reason is that, even in the $s\rightarrow \infty$ limit, the untruncated multi-particle fermionic Hilbert space keeps track of the odd-graded nature of $\gamma$. This is why, in that limit, parafermionic oscillators and not bosonic oscillators are recovered. It was already mentioned, at the beginning of this Section, that the
${\mathbb Z}_2$-grading of the Hilbert space implies a superselection of the states (bosons versus fermions) and that states of different gradings cannot be superposed. This superselection rule is absent in the purely bosonic theory.\par
~\par

This dual description (bosonic versus ${\mathbb Z}_2$-graded)  is a common feature in quantum models, even encountered in the quantization of the harmonic oscillator. The results of the famous \cite{wig} 1950 Wigner's paper  {"\it Do the Equations of Motion Determine the Quantum Mechanical Commutation Relations?}" can be reinterpreted
in modern language, see \cite{duality}, in this framework. Basically, in \cite{wig} Wigner flips the statistics of the $a^\dagger, a$ creation/annihilation operators of the harmonic oscillator, inducing an ${\mathfrak{osp}}(1|2)$ spectrum-generating superalgebra; he recovers the spectrum of the harmonic oscillator from a lowest weight rep of ${\mathfrak{osp}}(1|2)$, instead of the standard Fock-vacuum derivation. In 1950 Lie superalgebras had not yet been introduced in mathematics and, possibly,  Wigner's construction can be the first appearance of what is now called the ${\mathfrak{osp}}(1|2)$ superalgebra.\par 

In the ${\mathfrak{osp}}(1|2)$ derivation of the spectrum of the harmonic oscillator,  the ``bosonic" versus ``fermionic" labellings of the states refer to their respective parities under the $x\leftrightarrow -x$ parity transformation of the space coordinate;  the vacuum state (proportional to the gaussian $e^{-\frac{1}{2} x^2}$ for a convenient choice of parameters) is bosonic (even), the first excited state (proportional to $xe^{-\frac{1}{2} x^2}$) is fermionic (odd) and so on.

\section{Dynamical metasymmetries of the mixed-bracket Heisenberg-Lie algebras}

The mixed-bracket generalizations of the Heisenberg-Lie algebras introduced in Section {\bf 8} appear as dynamical symmetries
of Ordinary Differential Equations given by  Matrix Schr\" odinger equations in $0+1$ dimensions. This property is easily illustrated for the simplest nontrivial example which corresponds to the $2$-particle Hamiltonian $H_{2}$. The extension to  $N$-particle Hamiltonians with $N>2$ is immediate.\par
As recalled in Section {\bf 8}, the mixed-bracket Heisenberg-Lie algebras are not metaabelian according to the original \cite{lese1,lese2} definition; nevertheless,
they are generalized algebras which do not respect the even/odd ${\mathbb Z}_2$-grading of Lie superalgebras and which satisfy
an alternative metaabelianess condition given by the (\ref{newmeta}) identity. Acting as dynamical symmetries of an  ordinary Matrix Differential Equation, they deserve being referred to as {\it dynamical metasymmetries}.\par
The Matrix Schr\" odinger equation  under consideration is
\bea
\label{matrixschr}
\big(i\partial_t \cdot {\mathbb I}_4 - H_{2}\big)\Psi(t) &=& 0, 
\eea 

where $ H_{2}=diag(0,1,1,2)$ and $ \Psi(t)$ is a $4$-component vector.\par
The $\Psi_{ij}(t)$ solutions of (\ref{matrixschr}) can be expressed in terms of the creation $A_1^\dagger, A_2^\dagger$ and annihilation  
$A_1, A_2$ operators introduced in (\ref{braidedcr23},\ref{nonstandard2p}) and defined for the given angle $\pi g$. The further position $g=\frac{1}{s}$ for $s=2,3,4,5,\ldots$  defines the level-$s$ truncation. \par
We have
\bea
\Psi_{00}(t) &=& v_{00}, \qquad \qquad\qquad\qquad \qquad\qquad\quad~ {\textrm{where $v_{00}^T=(1,0,0,0)$,}}\nonumber\\
\Psi_{10}(t) &=& e^{-it} A_1^\dagger v_{00} = e^{-it}v_{10}, \qquad\qquad\qquad {\textrm{where $v_{10}^T=(0,0,1,0)$,}}\nonumber\\
\Psi_{01}(t) &=& e^{-it} A_2^\dagger v_{00} = e^{-it}v_{01}, \qquad\qquad\qquad {\textrm{where $v_{01}^T=(0,e^{i\pi g},0,0)$,}}\nonumber\\
\Psi_{11}(t) &=& e^{-2it} A_1^\dagger A_2^\dagger  v_{00} = e^{-2it}v_{11}, \qquad\qquad {\textrm{where $v_{11}^T=(0,0,0,e^{i\pi g})$.}}
\eea
Due to the commutators
\bea
\relax &[H_{2}, A_1^\dagger]= A_1^\dagger, \quad [H_{2}, A_2^\dagger]= A_2^\dagger,
\quad [H_{2}, A_1]=- A_1,\quad [H_{2}, A_2]= -A_2,&\nonumber\\
&&
\eea
 it is immediate that, by setting
\bea
&S_1^\dagger = e^{-it}A_1^\dagger,\qquad S_2^\dagger = e^{-it}A_2^\dagger,\qquad S_1 = e^{it}A_1,\qquad S_2 = e^{it}A_2,&
\eea
one ends up with four plus one symmetry operators (the extra operator being the $4\times 4$ identity operator $c:= {\mathbb I}_4$)  satisfying
\bea
\relax [S_\sharp, i\partial_t\cdot{\mathbb I}_4-H_{2}] &=& 0 \qquad \quad {\textrm{for $S_\sharp = S_1^\dagger,~ S_2^\dagger, ~S_1,~S_2,~ c$.}}
\eea
If $\Psi(t)$ is a solution of the matrix Schr\" odinger equation (\ref{matrixschr}),  then $\Psi'(t) = S_\sharp \Psi(t)$
is also (provided that $\Psi'(t)\neq 0$) a solution of the same equation.\par
~\par
Due to the fact  that the $e^{\pm i t}$ phases play no role as far as the brackets are concerned,  the five operators  $S_1^\dagger,~ S_2^\dagger, ~S_1,~S_2,~ c$ produce the same mixed-bracket Heisenberg-Lie algebra expressed by
the formulas  (\ref{g0bracket},\ref{anticommut},\ref{genuinemixed}). Up to the phases, the identification is
$c\equiv G_0, ~ S_1^\dagger \equiv G_{+1},~ S_1\equiv G_{-1},~ S_2^\dagger\equiv G_{+2}, ~S_2\equiv G_{-2}$. The simplest nontrivial case corresponds to the level-$3$ root of unity obtained from $g= \frac{1}{3}$. For this value
the $A_1^\dagger, A_2^\dagger$ matrices are those expressed in formulas (\ref{specialize},\ref{specialize2}).\par
~\par
A dynamical metasymmetry is also produced in the $s\rightarrow \infty $ limit; in this case the symmetry is given by the ${\mathbb Z}_2^2$-graded parafermionic oscillator algebra ${\mathfrak{h}}_{\mathfrak{pf}}(2)$ whose brackets are the (anti)commutators (\ref{paraf1},\ref{paraf2}). \par
It is worth pointing out that this is not the first example of a ${\mathbb Z}_2^2$-graded dynamical symmetry. It was shown in \cite{{aktt1},{aktt2}} that the symmetry operators of the Partial Differential Equations of the nonrelativistic L\'evy-Leblond spinors possess a ${\mathbb Z}_2^2$-graded Lie superalgebra structure. This can be regarded, in a different context, as another implementation of the ``symmetries wider than supersymmetry" concept.

\section{Conclusions}

Braided Majorana qubits have an obvious interest in the light of the Kitaev's proposal \cite{kit} to use them for encoding topological quantum computations which offer topological protection from quantum decoherence. Due to their neat combinatorics, the braided Majorana qubits are also an ideal playground to investigate different mathematical frameworks which can be applied to describe parastatistics. This is the content of the paper whose main results are here summarized.\par
~\par
The probably most relevant of such frameworks is based on the Leites-Serganova \cite{{lese1},{lese2}} notion of ``symmetries wider than supersymmetry", concerning the existence of statistics-changing maps which do not preserve the ${\mathbb Z}_2$-grading of ordinary Lie superalgebras. Possible applications to parastatistics (including the introduction of brackets which interpolate ordinary commutators and anticommutators) were suggested in the \cite{{lese1},{lese2}} mathematical papers. On the other hand this mathematical structure, 
as far as I know, mostly passed unnoticed in the physical literature. In this paper it is shown that the multi-particle sectors of braided Majorana qubits are described by a new class of algebras,  the generalized, mixed-bracket Heisenberg-Lie algebras introduced in Section {\bf 8}. \par
Rather unexpectedly, the original Leites-Serganova construction has to be relaxed to accommodate a more general setting. In \cite{{lese1},{lese2}} the statistics-changing maps were induced by nonhomogeneous subspaces of Lie superalgebras closed under the superbracket. This leads to the notion of {\it Volichenko algebras} which satisfy the (\ref{metabel}) metaabelianess condition (for any $X,Y,Z$ triple of operators, the identity
$ [X,[Y,Z]] = 0 $, which involves ordinary commutators, is satisfied).  The mixed-bracket Heisenberg-Lie algebras are not metaabelean, but satisfy the (\ref{newmeta}) identity, which can be referred to as a ``metaabelianess condition for the mixed brackets" (for any $X,Y,Z$ triple of operators, the $(X,(Y,Z))=0$ identity is satisfied, where $(.,.)$ denotes the mixed brackets). \par
The Leites-Serganova notion of {\it metasymmetry}, as discussed in Section {\bf 10}, can be applied to this more general setting. This is obviously quite a fascinating topic, both in pure mathematics and in connection with physics.
\par
~ \par  

Concerning the other results presented in this paper, let me first remind that the traditional framework to introduce parastatistics is based, see \cite{{gre},{grme}}, on the trilinear relations; more recently, the \cite{maj} approach based 
on graded Hopf algebras endowed with a braided tensor product was introduced (the connection between the two approaches has been investigated in \cite{{anpo1},{kada1}}).\par
~\par
The \cite{maj}  framework was applied in \cite{{top1},{top2}}  to prove the theoretical detectability of the paraparticles implied by the ${\mathbb Z}_2^2$-graded ``color" Lie (super)algebras introduced by Rittenberg-Wyler
in \cite{{riwy1},{riwy2}}. Later on, the extension to a higher dimensional representation of the braid group was applied
in \cite{topqubits} to first-quantize the multi-particle braided Majorana qubits. The truncations of the multi-particle spectra at roots of unity were observed. This obviously suggests, see \cite{{lus},{dck}}, a quantum group interpretation of the truncations. On the other hand, a quantum group interpretation is not directly available from the \cite{maj} framework applied to the ${\mathfrak{gl}}(1|1)$ superalgebra endowed with a braided tensor product. The puzzle is solved in Section {\bf 4}; the multi-particle spectra of the braided Majorana qubits are reconstructed from representations of the ${\cal U}_q({\mathfrak{osp}}(1|2))$ quantum supergoup. The next topic, from Section {\bf 5},
consists in  realizing (via the introduction of intertwining operators) the {\it braided tensor products} of the creation/annihilation operators in terms of ordinary tensor products.  This not only facilitates the computer-assisted algebraic manipulations (via Mathematica or any other favorite software); it lays the basis to construct the matrix representations which are needed in order to introduce the mixed-bracket Heisenberg-Lie algebras.\par
~\par
The last topic worth mentioning is the $s\rightarrow\infty$ {\it untruncated} limit of the mixed-bracket 
Heisenberg-Lie algebras presented in Section {\bf 9}. The algebraic structures recovered in the limit are (once more) the Rittenberg-Wyler color Lie superalgebras. In this limit the brackets are no longer interpolations of commutators/anticommutators, but they are reduced to ordinary (anti)commutators. Nevertheless, they are organized in a different way with respect to the ordinary (anti)commutators of the bosons/fermions statistics, see e.g. the entries in the (\ref{z2z2parastat}) table. The $s\rightarrow \infty$ limit implies paraparticles obeying the types of parastatistics investigated in  \cite{{top1},{top2}}.\par
~\par
As a final remark, analyzing in retrospect the results of \cite{{aktt1},{aktt2}} (the ${\mathbb Z}_2^2$-graded Lie superalgebra structure of the symmetry operators of the nonrelativistic L\'evy-Leblond spinors), one can conclude (see the remark at the end of Section {\bf 10}) that this is another example of the Leites-Serganova notion of  ``symmetries wider than supersymmetry".  \par
The results of this paper point towards a more general setting than the one investigated in \cite{{lese1},{lese2}} to accommodate statistics-changing maps; it can be applied to braid statistics recovered from quantum groups, parastatistics recovered from color Lie (super)algebras, etc.

 \par
~\par
~\par
\newpage

{\Large{\bf{Appendix A: a comment on mixed brackets and quons}}}
\par
~\par
It is worth mentioning that ``mixed brackets" which interpolate commutators and anticommutators can be defined for other types of parastatistics. Perhaps the most notable example is the algebra of quons which was introduced by Greenberg \cite{{gre1,{gre2}}} and Mohapatra \cite{moh}. Quons are $q$-deformed oscillators, defined for $-1\leq q \leq 1$ (the constraint expressing the admissible quonic Hilbert spaces is discussed in \cite{gre2}) which interpolate between fermions ($q=-1$) and bosons ($q=1$).
The ``$q$-mutators" of $n$ creation/annihilation  $a_i^\dagger, a_i$ quons, with $i=1,2,\ldots, n$ are defined to satisfy  
\bea\label{qmutator}
a_ia_j^\dagger -q a_i^\dagger a_j &=& \delta_{ij}.
\eea
Since its introduction, the quonic parastatistics has been intensely investigated; one can mention, e.g., the paper
\cite{bdm} relating quons to anyons.\par
It is a trivial exercise to express the $q$-mutator (\ref{qmutator}) of one ($n=1$) quon as a mixed bracket,
interpolating commutator and anticommutator. One has to set
\bea\label{onequon}
aa^\dagger -q a^\dagger a= 1 &~~\Leftrightarrow ~~& \cos^2(\theta_q)\cdot[a,a^\dagger] + \sin^2(\theta_q)\cdot\{a,a^\dagger\} =1,
\eea
where the angle $\theta_q$, comprised in the range $\theta_q\in [0,\pi/2]$, is given by
\bea\label{quonangle}
\theta_q &=&\arcsin \left(\sqrt{\frac{1-q}{2}}\right), \qquad (q= 1-2\sin^2\theta_q).
\eea
Special $\theta_q$ angles are\\
~\\
$~~~ i$)  ~~ $\theta_q=0 ~$~  for $q=1$, which recovers the bosons,\\
$~~i i$) ~~ $\theta_q=\frac{\pi}{4} ~$  for $q=0$, which recovers the so-called, see \cite{gre1}, ``infinite statistics" case,\\
$iii$) ~~ $\theta_q=\frac{\pi}{2} ~$  for $q=-1$, which recovers the fermions.\\

The mixed-bracket Heisenberg-Lie algebras, which have been introduced in Section {\bf 8} and reproduce the multi-particle parastatistics of the braided Majorana qubits,
differ from the quonic ``mixed brackets" as derived in formulas (\ref{onequon},\ref{quonangle}).
Indeed, the (\ref{newbuildingn}) building blocks $A_{k;N}^\dagger$ of the creation operators of the $N$-particle sector and their  $A_{k;N} =(A_{k;N}^\dagger)^\dagger$ annihilation counterparts do not satisfy  the quonic algebra (\ref{onequon}) for any nontrivial $q\neq \pm 1$ value; it is easily shown that they satisfy, for any given $k=1,2,\ldots, N$, an ordinary anticommutator:
\bea
A_{k;N}A_{k;N}^\dagger +A_{k;N}^\dagger A_{k;N} &=&{\mathbb I}_{2^N}.
\eea
The genuine mixed brackets $(\cdot,\cdot)_\vartheta$ from (\ref{volimixedbracket}) with a non-trivial angle dependence ($\vartheta\neq n\frac{\pi}{2}$ for integer values of $n$)  are taken with
respect to the creation/annihilation operators for $k\neq k'$, as shown, e.g.,  in formulas  (\ref{genuinemixed}) for the $2$-particle sector. At level-$s$ ($s=2,3,4,\ldots$) we get, in particular,
\bea
&(G_{+1},G_{+2})_{+\frac{s+2}{2s}\pi}\equiv (A_{1;2}^\dagger, A_{2;2}^\dagger )_{+\frac{s+2}{2s}} = 0.
&
\eea
It should be stressed, as a final comment, that the parastatistics induced by the mixed-bracket Heisenberg-Lie algebras are not related to the quonic parastatistics. 

~\par
~\par

{\Large{\bf{Appendix B: about third-roots of unity and ternary algebras}}}
\par
~\par
In the minimal matrix representation, the $N$-particle sector of the  braided Majorana qubits is realized, see ({\ref{newbuildingn}), by $2^N\times 2^N$ matrices. Equivalent descriptions which produce isomorphic Hilbert spaces can be obtained from nonminimal representations. Let's consider the third root of unity; an example of a set of nonminimal representations is given
by $2\cdot 6^{N-1}\times 2\cdot 6^{N-1}$ matrices. Unlike the minimal representations (\ref{newbuildingn}) with the special third root of unity case in (\ref{specialize}, \ref{specialize2}), this nonminimal set encodes a ${\mathbb Z}_3$ ternary grading. It can therefore finds application to ternary physics, see for details \cite{akl} and references therein (one of such cases is the  fractional supersymmetry discussed in \cite{rtsi}). 
Despite providing the same description of braided Majorana fermions, the minimal representation (\ref{newbuildingn}) at the third root of unity does not respect the ${\mathbb Z}_3$ grading and, therefore, cannot be
used in these ternary physics applications.\par
The ternary construction of the braided Majorana qubits employs tensor products of the three $3\times 3$ matrices
 $Q_i$
(defined for $j=e^{i{\frac{2}{3}\pi}}$ with $j^3=1$) and their $Q_i^\dagger$ hermitian conjugates:
{\small{
\bea\label{ternary}
&Q_1=\left(\begin{array}{ccc}0&1&0\\0&0&j\\j^2&0&0\end{array}\right),\qquad 
Q_2=\left(\begin{array}{ccc}0&j&0\\0&0&1\\j^2&0&0\end{array}\right), \qquad
Q_3=\left(\begin{array}{ccc}0&1&0\\0&0&1\\1&0&0\end{array}\right),
 &\nonumber\\
&Q_1^\dagger=\left(\begin{array}{ccc}0&0&j\\1&0&0\\0&j^2&0\end{array}\right),\qquad 
Q_2^\dagger=\left(\begin{array}{ccc}0&0&j\\j^2&0&0\\0&1&0\end{array}\right), \qquad
Q_3^\dagger=\left(\begin{array}{ccc}0&0&1\\1&0&0\\0&1&0\end{array}\right).
 &
\eea
}}
Following \cite{akl} a consistent ${\mathbb Z}_3$ grading can be assigned by setting, $mod~ 3$,
\bea
&
deg (Q_i)=1, \qquad deg (Q_i^\dagger) = 2,\qquad {\textrm{for $i=1,2,3$.}}&
\eea
The non-minimal building blocks of the braided $2$-particle creation/annihilation operators are
\bea\label{2pternary}
&{\widetilde{A}_1}^\dagger =\gamma\otimes {\mathbb I}_2\otimes Q_1,\quad {\widetilde {A}_2}^\dagger =\gamma\otimes {\mathbb I}_2\otimes Q_2,\quad {\widetilde {A}_1} =\gamma^\dagger\otimes {\mathbb I}_2\otimes Q_1^\dagger,\quad {\widetilde {A}_2}=\gamma^\dagger\otimes {\mathbb I}_2\otimes Q_2^\dagger,&
\eea
while for the $3$-particle sector we have {\small{
\bea\label{3pternary}
&{\widetilde {B}_1}^\dagger =\gamma\otimes {\mathbb I}_2\otimes {\mathbb I}_2\otimes Q_1\otimes  {\mathbb I}_3,\quad 
{\widetilde B_2}^\dagger ={\mathbb I}_2\otimes\gamma\otimes {\mathbb I}_2\otimes Q_2\otimes Q_1,\quad{\widetilde B_3}^\dagger ={\mathbb I}_2\otimes{\mathbb I}_2\otimes\gamma\otimes Q_2\otimes Q_2.&
\eea
}}
The nilpotent matrix $\gamma$, introduced in (\ref{4op}), is 
{\footnotesize{$\gamma =\left(\begin{array}{cc} 0&0\\1&0\end{array}\right)$}}.\par
The ternary $2$-particle representation  (\ref{2pternary}) is given by $12\times 12$ matrices. The $3$-particle ternary representation (\ref{3pternary}) with operators ${B}_i^\dagger$ and their hermitian conjugates is
given by $72\times 72$ matrices.  The extension of the construction to the $N>3$ particle sectors, producing  $2\cdot 6^{N-1}\times 2\cdot 6^{N-1}$ matrices, is immediate.
\par
The nilpotency of $\gamma$ implies
\bea\label{newnil2}
 ({\widetilde A_i}^\dagger)^2=0 \quad{\textrm{for $i=1,2$}} \qquad {\textrm{and}}\qquad 
 ({\widetilde B_j}^\dagger)^2=0 \quad{\textrm{for $j=1,2,3$.}}
\eea
We further get the braiding relations
\bea\label{2pternarybraid}
({\widetilde {A}_2}^\dagger\cdot {\widetilde {A}_1}^\dagger)& =& j\cdot ({\widetilde{A}_1}^\dagger  \cdot{\widetilde{A}_2}^\dagger) \qquad\qquad {\textrm{and}}
\eea
{\small{
\bea\label{3pternarybraid}
&({{\widetilde B}_2}^\dagger  {\widetilde{B}_1}^\dagger) = j\cdot ({\widetilde {B}_1}^\dagger {\widetilde {B}_2}^\dagger), \quad
({\widetilde B_3}^\dagger{\widetilde  B_1}^\dagger) = j\cdot ({\widetilde{B}_1}^\dagger {\widetilde B_3}^\dagger), \quad
({\widetilde B_3}^\dagger {\widetilde {B_2}}^\dagger) = j\cdot ({\widetilde B_2}^\dagger  {\widetilde{B}_3}^\dagger).&
\eea
}}
Formulas (\ref{newnil2},\ref{2pternarybraid},\ref{3pternarybraid}) reproduce, for $g=\frac{1}{3}$, the corresponding
``minimal" formulas presented in (\ref{newnil},\ref{newbraid2p},\ref{newbraid3p}).\par
In the non-minimal setting the quantization of the braided Majorana qubits at the third root of unity is reproduced
by introducing the $N$-particle vacuum $|{\widetilde{ vac}}\rangle_N$ as the $2\cdot 6^{N-1}$ column vector
$|{\widetilde{ vac}}\rangle_N=e_1$, where $e_1$ denotes the column with first entry $1$ and $0$ otherwise.\par
For $N=2,3$ the Hamiltonians ${\widetilde{H_2}}$, ${\widetilde {H_3}}$ are respectively given by
\bea
{\widetilde{H_2}}&=&(\delta\otimes {\mathbb I}_2+{\mathbb I}_2\otimes\delta)\otimes {\mathbb I}_3,\nonumber\\
{\widetilde{H_3}}&=& (\delta\otimes {\mathbb I}_2\otimes{\mathbb I}_2+{\mathbb I}_2\otimes\delta\otimes{\mathbb I}_2+{\mathbb I}_2\otimes{\mathbb I}_2\otimes\delta)\otimes {\mathbb I}_3\otimes{\mathbb I}_3.
\eea
By introducing
\bea
{\widetilde A}^\dagger = {\widetilde {A}}_1^\dagger+{\widetilde {A}}_2^\dagger,&\quad &{\widetilde B}^\dagger = {\widetilde {B}}_1^\dagger+{\widetilde {B}}_2^\dagger +{\widetilde {B}}_3^\dagger,
\eea
the $2,3$-particle Hilbert spaces ${\widetilde{\cal H}}_{(2)}$, ${\widetilde{\cal H}}_{(3)}$ are respectively spanned
by the vectors
\bea
({\widetilde A}^\dagger)^n |{\widetilde{ vac}}\rangle_2\in {\widetilde{\cal H}}_{(2)},&& 
({\widetilde B}^\dagger)^n |{\widetilde{ vac}}\rangle_3\in {\widetilde{\cal H}}_{(3)}, \qquad {\textrm{ for $n=0,1,2,\ldots $.}}
\eea
The truncation at the third root of unity is recovered from the identity
\bea
({\widetilde B}^\dagger)^3&=&0,
\eea
which implies that the $3$-particle sector is spanned by the three energy eigenvectors $|{\widetilde{ vac}}\rangle_3$, ${\widetilde B}^\dagger|{\widetilde{ vac}}\rangle_3$ and $({\widetilde B}^\dagger)^2 |{\widetilde{ vac}}\rangle_3$ whose respective energy eigenvalues are $E=0,1,2$.\par
~\par
The matrices in (\ref{2pternary},\ref{3pternary}) reproduce the mixed brackets of, respectively, the $2$-particle and $3$-particle Heisenberg-Lie algebras at the level $s=3$  (that is, the third root of unity). They induce the same algebras as the ones recovered from the corresponding minimal representations introduced in Section {\bf 5} (see, e.g., formula  (\ref{braidedcr23}) specialized at the  $g=\frac{1}{3}$ value).

\newpage
 {\Large{\bf Acknowledgments}}
{}~\par{}~\\
Many years ago during a visit in Stockholm Dimitry Leites illustrated and  strongly stressed to me the relevance of Volichenko algebras (at that time I could not find a proper application for this structure). More recently, at a Conference in Troms\o,  I discussed a preliminary version of this paper with Vera Serganova (I was not yet aware that,
for braided Majorana qubits, the metaabelianess condition has to be relaxed). Concerning the puzzling feature of the roots of unity truncations obtained in the previous work on braided Majorana qubits, I had a few discussions and exchanges of ideas with Nikolai Reshetikhin and Calvin McPhail-Snyder.   The quantum group rep interpretation of the truncations, presented in Section {\bf 4}, benefited from these discussions. The ${\mathbb Z}_2^2$-graded parafermionic oscillators recovered in the $s\rightarrow +\infty$ untruncated limit are among the structures we constantly discuss with Naruhiko Aizawa and Zhanna Kuznetsova. \\
 This work was supported by CNPq (PQ grant 308846/2021-4).

\end{document}